\documentclass[useAMS,usenatbib]{mn2e}

\usepackage{amsmath,texlogos}
\usepackage{natbib}
\usepackage{subfigure}
\usepackage{graphicx}
\usepackage{txfonts}
\usepackage[T1]{fontenc}
\usepackage{aecompl}
\usepackage{times}

\title[DOHA Algorithm]{The DOHA algorithm: a new recipe for cotrending large-scale transiting exoplanet survey light curves.}
\author[Mislis D.]{Mislis D.$^{1}$\thanks{E-mail:dmislis@qf.org.qa}, Pyrzas S.$^{1}$, Alsubai K.A.$^{1}$, Tsvetanov Z.I. $^{1}$, Vilchez N.P.E. $^{1}$ \\ \\
$^{1}$Qatar Environment and Energy Research Institute, LAS Building, Education City, Qatar Foundation, P.O. Box 5825, Doha, Qatar\\
}

\begin{document}

\date{Accepted ??. Received ??; in original form ??}

\pagerange{\pageref{firstpage}--\pageref{lastpage}} \pubyear{2002}

\maketitle

\label{firstpage}

\begin{abstract}
We present \texttt{DOHA}, a new algorithm for cotrending photometric light curves obtained by transiting exoplanet surveys. The algorithm employs a novel approach to the traditional
``differential photometry'' technique, by selecting the most suitable comparison star for each target light curve, using a two-step correlation search. Extensive tests on real data
reveal that \texttt{DOHA} corrects both intra-night variations and long-term systematics affecting the data. Statistical studies conducted on a sample of $\sim$9\,500 light curves 
from the Qatar Exoplanet Survey reveal that \texttt{DOHA}-corrected light curves show an RMS improvement of a factor of $\sim\,2$, compared to the raw light curves. In 
addition, we show that the transit detection probability in our sample can increase considerably, even up to a factor of 7, after applying \texttt{DOHA}.

\end{abstract}


\section{Introduction}

In the last decade, a significant portion of the hunt for transiting extrasolar planets has been conducted by various ground-based, large-scale surveys, such as SuperWASP \citep{swasp}, 
HatNet \citep{hatn}, TrES \citep{tres} and QES \citep{alsubai}. A common, defining characteristic of these surveys is that they were designed to cover as large a field of view as 
possible. 

Data obtained by these surveys tend to suffer from a, more or less, common problem: the presence of unwanted flux variations that can either mask or mimic real (astrophysical)
variations. A significant part of these variations is introduced by fixed, ordered trends in the data, collectively referred to as ``systematics''. The list of systematics is rather 
long including, among others, variations due to airmass and seeing, colour-dependent extinction, object merging etc. The imprint of systematics on the data can be viewed as components 
leading to common-mode behaviour among the light curves of observed stars. 

In addition, unwanted flux variations can also be introduced by random events. By definition these are not systematic, or in other words, they are events that do not have a 
distinct, common mode imprint on the data (see e.g. \citet{pinheiro}).

As the photometric accuracy for ground-based exoplanet detection is required to be of the order of 1\% or better, it became readily apparent that all these variations, with
amplitudes that often exceed a few percent and ``signatures'' that can easily mimic a transit event, can severely reduce the transit detection probability and, therefore, need to be 
accounted for and corrected. This has lead to detrending algorithms such as \texttt{TFA} \citep{kovacs2} and \texttt{SysRem} \citep{tamuz}. Recently, similar work has been 
done for space missions such as CoRoT \citep{mislis2, ofir} and Kepler \citep{still}.

While they differ in their implementation, the core idea of these algorithms remains the same: they try to identify and correct systematic patterns, by exploiting their common 
mode behaviour. A crucial factor in this exercise is the actual commonness of the patterns, in the (statistical) sense of what percentage of stars are affected by them, or in other 
words, how representative the patterns are of the entire sample. An additional consideration is the quantitative contribution of each pattern on the overall variations and whether 
specific patterns can be viewed as driving the variations. We maintain the distinction between common and uncommon dominant patterns throughout the manuscript.

In this paper we present \texttt{DOHA}, an algorithm conceived to correct for both systematic variations (regardless of commonness) and assorted data irregularities. The structure of 
the paper is as follows: in Section\,\ref{sec:sample} we briefly describe the data set used in testing the algorithm; in Section\,\ref{sec:algor} we present the algorithm itself, 
while Section\,\ref{sec:resu} contains the results after applying \texttt{DOHA} to our sample light curves. Section\,\ref{sec:sigdet} shows a test for signal detection efficiency and 
Section\,\ref{sec:conc} summarises our work.

\begin{figure*}
 \includegraphics[width=\textwidth]{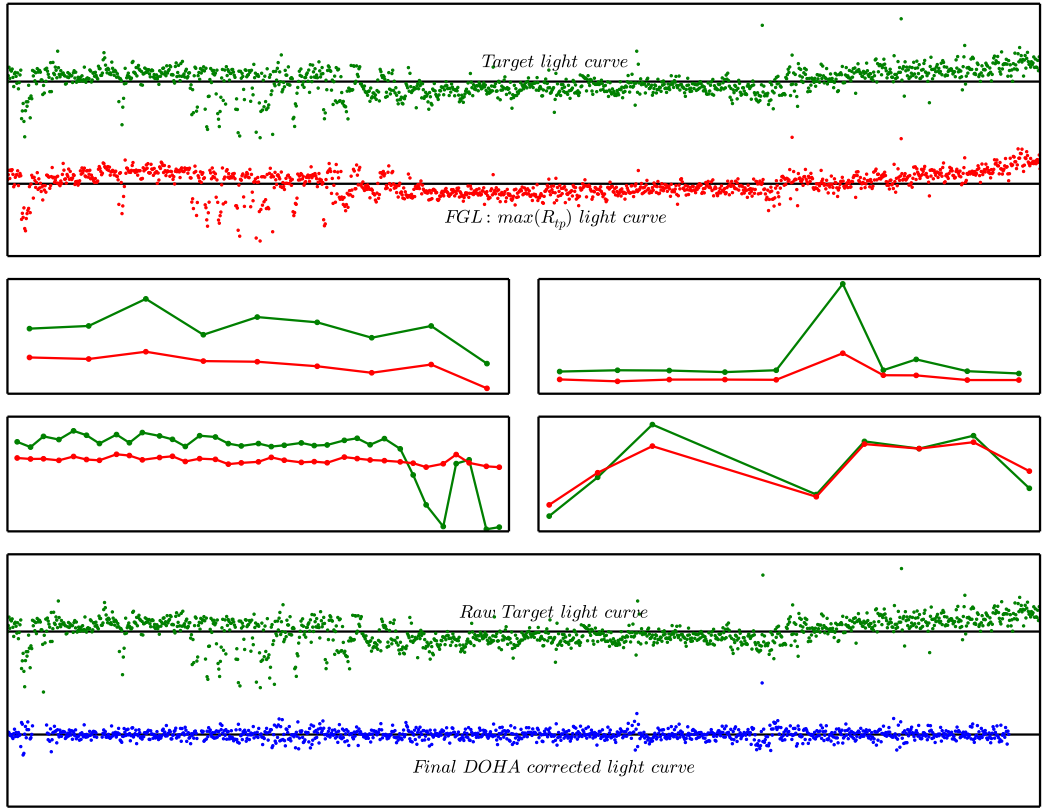}
 \caption{\textbf{Top:} raw target light curve (green), and light curve of the \emph{highest correlated} comparison star from the FGL (red). For clarity, light curves are plotted as 
 consecutive points and \emph{not} according to their timestamps. \textbf{Middle:} raw target light curve (green) and master comparison light curve (red) for four individual nights. 
 \textbf{Bottom:} raw target light curve (green) and final, \texttt{DOHA}-corrected light curve (blue). Note: light curves have been normalized to unity by their respective mean 
 flux. See text for additional details.}
 \label{fig1}
\end{figure*}

\section{The sample data}
\label{sec:sample}
For testing our algorithm we used data from the Qatar Exoplanet Survey (QES). The QES uses six 4k$\times$4k FLI ProLine PL6801 cameras, equipped with 4x400mm, 1x200mm and 1x135mm 
lenses, mosaiced to image an $11^{o}\,\times\,11^{o}$ field on the sky in the magnitude range of $10 < m_{V} < 17$. For our purposes, we selected data from a single field 
(RA=3$^{h}$50$^{m}$, DEC=-3$^{o}$00'), obtained with one of the cameras equipped with a 400mm lens (FOV=$5.24^{o}\times5.24^{o}$). The data were collected over a three-month 
period (end Mar-Jun 2010) and consist of $\sim$9\,500 stars, with an average of $\sim$1\,300 data points each and an exposure time of 60 sec. The data were reduced using the 
QES pipeline as described in \cite{alsubai}.

\section{The algorithm}
\label{sec:algor}
Let us assume that our data set consists of light curves $f\left(n\right)$, with $n$ number of data points each, and that the total number of light curves is $N$. We wish to correct 
the light curve of the $t$-th \emph{target} star $f_{t}\left(n\right)$. \texttt{DOHA} achieves correction using a two-step correlation search approach.

In the first step, the algorithm calculates the correlation coefficient $R_{tp}$ between the target star $f_{t}\left(n\right)$ and a \emph{potential} comparison star 
$f_{p}\left(n\right)$. Only stars with $\rm{RMS}^{P}\,<\,1.5*\rm{RMS}^{T}$ are considered as potential comparison stars, with $\rm{RMS}^{P}$ and $\rm{RMS}^{T}$ the RMS 
values of the raw comparison and target light curves respectively. The correlation coefficient is calculated as

\begin{equation}
\centering
R_{tp} = \frac{1}{n-1}\sum_{i=1}^{n} \frac{(f_{t}(i)-\overline{f_{t}}) \cdot (f_{p}(i)-\overline{f_{p}})}{ \sigma_{f_{t}} \cdot \sigma_{f_{p}}},
\end{equation}
\noindent
where $f_{t}(i)$ and $f_{p}(i)$ are the $i$-th data points of the target and potential comparison star light curves respectively; $\overline{f_{t}}$ and $\overline{f_{p}}$ are the 
mean flux values of the light curves; and $\sigma_{f_{t}}$ and $\sigma_{f_{p}}$ are the standard deviations of the target and comparison light curves respectively. We note
that $R_{tp}$ is calculated on the \emph{common set of points} of $f_{t}\left(n\right)$ and $f_{p}\left(n\right)$; missing points are \emph{not} substituted.

At the end of this first step, $N^{*}$ values of $R_{tp}$ correlation coefficients have been calculated in total. Subsequently, the mean and standard deviation of all $R_{tp}$ 
values are derived, and those stars that have corresponding $R_{tp}$ values larger than 2-$\sigma$ from the mean are selected. In this fashion, we create a ``family'' of 
\emph{definite} comparison stars, of size $D$, with light curves $f_{d}(n)$, $d\,=\,1,2,\ldots,D$. Henceforth, we will denote this as \textit{Family Group Light curves} (FGL).

We should note that the 2-$\sigma$ cutoff limit is not cast in stone. It is a ``middle-ground'' balance between selecting an adequate number of stars for the FGL on the one hand; and 
selecting only those stars that are strongly correlated with the target on the other. The limit can be adjusted to better suit the given data set, depending e.g. on the number of stars 
with sufficient data points and on the severity of the systematic and non-systematic trends.

In the second step, the algorithm splits the light curve of the target star $f_{t}(n)$ to its individual-night segments. Working in each segment separately, the algorithm re-calculates
$R_{td}$ correlation coefficients, but this time, only the $f_{d}(n)$ light curves from the FGL are taken into consideration. We use single-night segments to better account for airmass 
and colour-extinction variations. 

As before, the mean and standard deviation of the values of $R_{td}$ are calculated and those stars with $R_{td}$ larger than 1.5-$\sigma$ from the mean are selected to 
create the \textit{New Family Group Light curves} (NFGL). Assuming that the NFGL is of size M, we will denote its light curves as $f_{m}(n)$, $m\,=\,1,2,\ldots,M$. We 
re-iterate that the NFGL is created for \emph{each} individual night segment; there are as many NFGLs as there are individual nights in the data. Also note that the constituent
light curves $f_{m}(n)$ of one NFGL are not necessarily the same as those of an other.

Again, the 1.5-$\sigma$ cutoff is a reasonable ``default'' value, adjusted to account for the much smaller-sized FGL (compared to the original number of light curves), following the 
considerations described previously.

Still working on an individual night basis (inb), from the corresponding NFGL, we create a "master" comparison light curve $f_{C,inb}(n)$, which is the mean of all the 
$f_{m}(n)$ light curves in the given NFGL,
\begin{equation}
\centering
f_{C,inb}(n) = \frac{1}{M} \sum_{m=1}^{M} f_{m}(n)
\end{equation}
\noindent
Once the master comparison curve is calculated, correction of the target light curve $f_{t,inb}$ (for the given individual night) is achieved via a double-iterative, 
``global'' RMS minimisation technique, as follows:

\begin{itemize}
 \item[-] We construct an array of \emph{scaling correction factors} $S_{i}$, of size $I$ ($i=1,\ldots,I$), with $S_{i}\in\left(0,1\right)$ and arbitrary step; for our tests, we chose 
 a step of 0.01\footnote{In which case, I=99; $S_{1}=0.01$ and $S_{I}=0.99$}
 \item[-] We define RMS$^{\,0}$, the RMS of the raw target light curve segment, $f^{\,0}_{t,inb}$, as the reference, starting point
 \item[-] An I-steps iteration over all $S_i$ begins 
 \item[-] Subsequently, a $j$-steps iteration begins, $j=1,2,\ldots$. For the given $S_i$, at the $j$-th step, a \emph{temporary} corrected target light curve ${f^{j}_{t,inb}}_{(i)}$ 
 is constructed by
\begin{equation}
 \centering
  {f^{j}_{t,inb}}_{(i)}\,=\,{f^{j-1}_{t,inb}}_{(i)}\,-\,S_{i} \cdot f_{C,inb}
\label{eq1}
\end{equation}
 \item[-] The RMS of this light curve, ${\mathrm{RMS}^{j}}_{(i)}$, is calculated and compared to that of the previous $j$-step, ${\mathrm{RMS}^{j-1}}_{(i)}$
 \item[-] Iterations over $j$ halt when ${\mathrm{RMS}^{j}}_{(i)}\,\geq\,{\mathrm{RMS}^{j-1}}_{(i)}$
 \item[-] The I-steps iteration continues with $S_{i+1}$, and the entire process is repeated
\end{itemize}

\noindent
The corrected target light curve segment $f^{\,\mathrm{cor}}_{t,inb}$ is chosen to be the one with the minimum RMS, that is 
$\mathrm{RMS}^{\mathrm{cor}}\,=\,\mathrm{MIN}\left({\mathrm{RMS}^{j}}_{(i)}\right)$. The \emph{final} \texttt{DOHA}-corrected light curve of the target star 
$f^{\,\mathrm{fin}}_{t}$ is simply the concatenation of all corrected target light curve segments $f^{\,\mathrm{cor}}_{t,inb}$.	

Finally, we note that, as with the $\sigma$-cutoffs applied in the creation of the FGL and the NFGL, the array-step for the $S_{i}$ values can be adjusted to better suit a given
data set.

\begin{figure}
 \includegraphics[width=9cm]{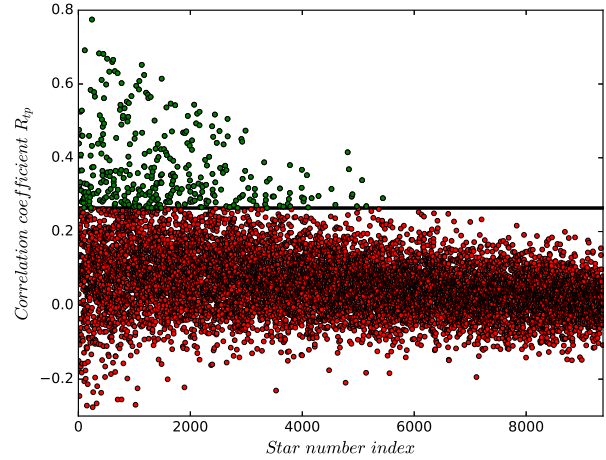}
 \caption{Correlation diagram for a randomly selected target star. The black, solid line indicates our 2-$\sigma$ selection criterion. The (green) points above this line form the FGL.}
 \label{fig2}
\end{figure}

For additional clarity, we provide representative illustrations of the algorithm's description, as presented above, in Figures\,\ref{fig1} and \ref{fig2}.

In the top panel of Figure\,\ref{fig1} we plot a (randomly selected) raw target light curve $f^{\,0}_{t}$ (green curve), and the light curve of the \emph{highest correlated} comparison 
star from the FGL (red curve). To better highlight variations around the respective mean values (indicated as solid lines), light curves are plotted as consecutive points and 
\emph{not} according to their timestamps, which span a range of three months. In the four middle-panels, we show four different individual night segments and we plot the corresponding 
$f^{\,0}_{t,inb}$ points (again in green), along with the master comparison light curve $f_{C,inb}$ (in red) for that particular night. Finally, in the bottom panel we plot again the 
raw target light curve $f^{\,0}_{t}$ (green) as well as the final, \texttt{DOHA}-corrected light curve $f^{\,\mathrm{fin}}_{t}$ (blue).

Note that \texttt{DOHA} does not apply explicit outlier rejection. Most of the outlying points in the raw light curve (Fig.\,\ref{fig1}, bottom panel, top curve) were actually 
``brought-in-line'' by the algorithm's correction.

Figure\,\ref{fig2} shows the $N^{*}$ values of the $R_{tp}$ correlation coefficients, calculated through Equation\,\ref{fig1}, for a random target star. The black, solid line 
corresponds to the 2-$\sigma$ level away from the mean. All the stars above this line (green points) are used to form the FGL. 

Note that our data are magnitude-sorted (brightest to faintest). The target star used in Fig.2 is a 11.3-mag one. Given the $\rm{RMS}^{P}\,<\,1.5*\rm{RMS}^{T}$ criterion and the fact the bright stars are more likely
to have high-correlation comparisons (this is explained in detail in the next Section), it is not surprising that most of the FGL members clump on the left-hand side of 
Figure\,\ref{fig2}. i.e. have a small star-index.

\section{Results}
\label{sec:resu}

\subsection{Statistical tests}
\label{stats}
\begin{figure}
 \includegraphics[width=9cm]{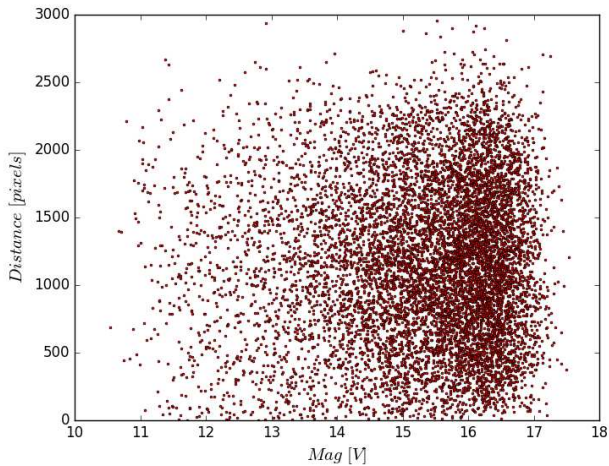}
 \caption{Target magnitude versus distance (in pixels) of the highest correlated comparison star for the given target. The distribution is totally random, and highest correlated 
 stars can be found anywhere on chip, independent of target magnitude.}
 \label{fig3}
\end{figure}

For each of the $\sim$9\,500 stars in our field, we constructed the corresponding FGL, noted the on-chip position of the target and of the highest correlated star in the FGL and
calculated their distance in pixels. In Figure\,\ref{fig3} we plot the target magnitude versus the distance to the highest correlated comparison star. The distance is given in 
pixels, with 1 pixel corresponding to $4.64^{''}$. Figure\,\ref{fig3} does not show any distinct pattern between the two plotted quantities. Effectively, the highest correlated star 
for any given target can be located anywhere on the chip.

\begin{figure}
\includegraphics[width=9cm]{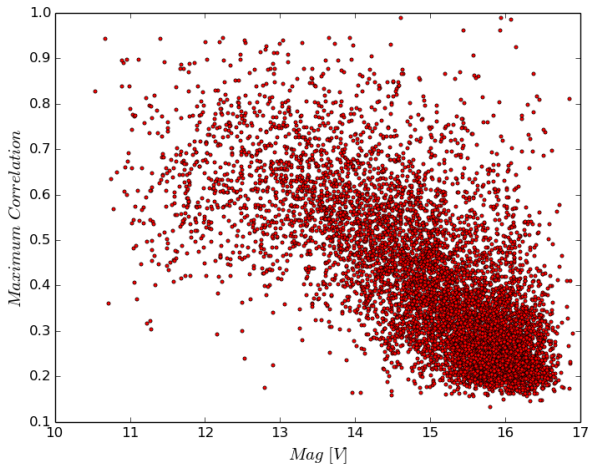}
 \caption{Target magnitude versus $\mathrm{MAX}\left(R_{tp}\right)$ correlation coefficient value. A linear trend is clearly visible, and comparison stars for the brighter targets
 show higher $R_{tp}$ values.}
 \label{fig4}
\end{figure}

Additionally, we noted the actual $R_{tp}$ value of the highest correlated comparison for a given target and in Figure\,\ref{fig4} we plot these $\mathrm{MAX}\left(R_{tp}\right)$ 
against target magnitude. Despite the scatter, a linear trend is clearly visible in Figure\,\ref{fig4}, showing that for bright stars, it is much more likely to find a comparison star with a high $R_{tp}$ 
value. This could be an indication that red noise dominates the systematics, and bright stars are more susceptible to red noise, as they tend to occupy a larger area (more pixels)
on the CCD.

\begin{figure}
  \includegraphics[width=9cm]{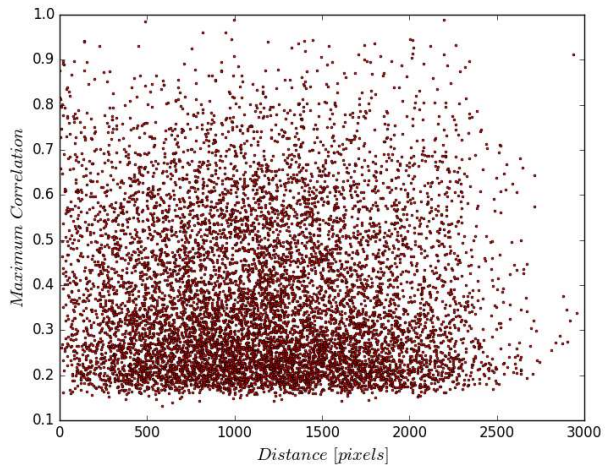} 
 \caption{Distance of the highest correlated comparison star versus maximum correlation value. Once more, there is no obvious trend between the two quantities; even for very 
 high $R_{tp}$ values (>0.9), the corresponding comparison stars can be located anywhere on the chip.}
 \label{fig3b}
\end{figure}

Finally, we combine Figures\,\ref{fig3} and \ref{fig4} and plot the distance of the highest correlated comparison star versus the corresponding $R_{tp}$ value in 
Figure\,\ref{fig3b}. As with Fig.\,\ref{fig3}, there is again no distinct pattern between the two plotted quantities. The value of maximum correlation is independent of the distance
between target and comparison, even for high $R_{tp}$ values (>0.9).

\begin{figure*}
\includegraphics[width=5.8cm]{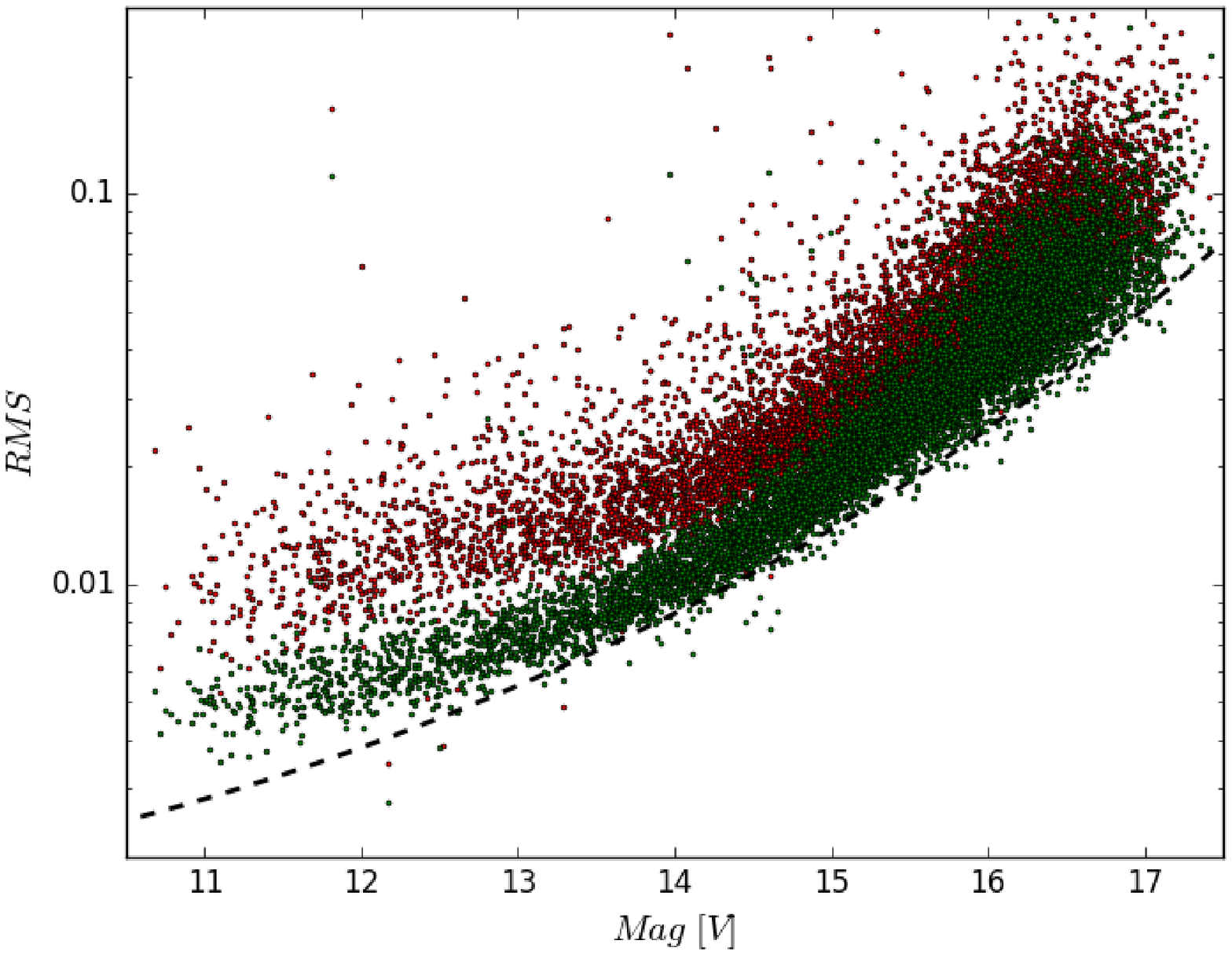}
\includegraphics[width=5.8cm]{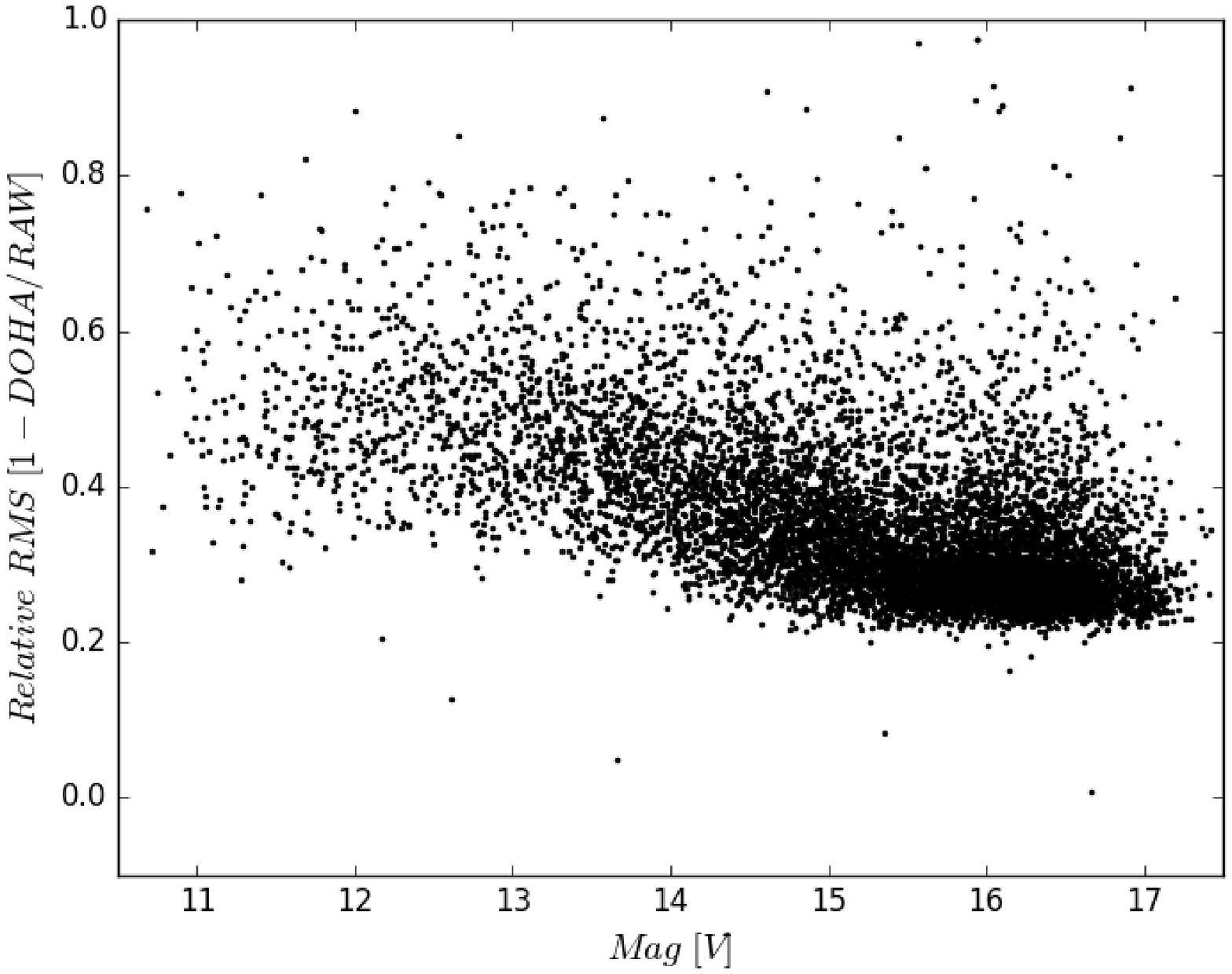} 
\includegraphics[width=5.8cm]{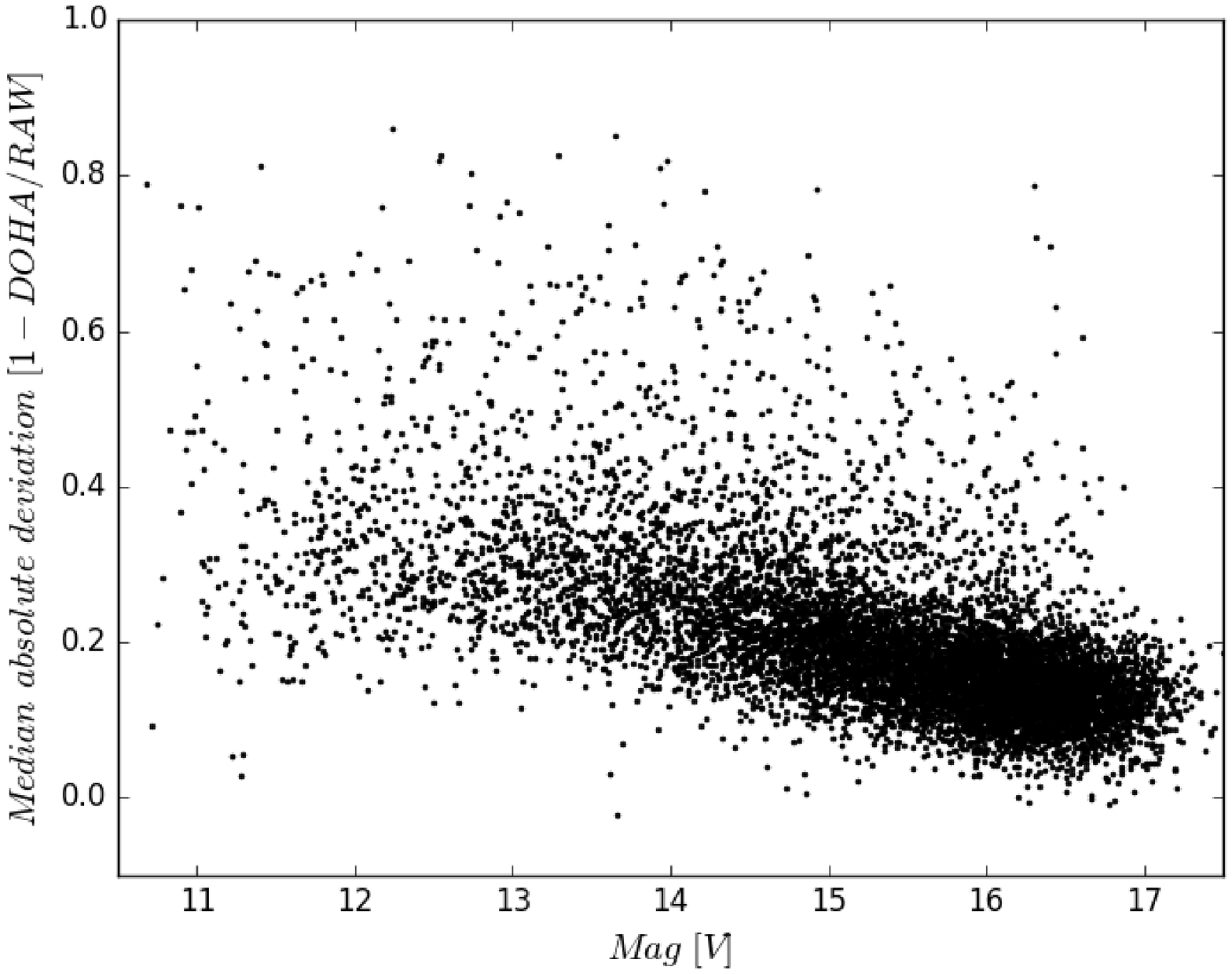} 
 \caption{\textbf{Left:} The RMS diagram of the sample before (red dots), and after \texttt{DOHA} (green dots). The dashed black line indicates the theoretical noise floor
 curve. \textbf{Middle:} Relative improvement between raw data and \texttt{DOHA}-corrected data versus target magnitude, using RMS-statistics. \textbf{Right:} Relative improvement
 using MAD-statistics.}
 \label{fig5}
\end{figure*}

\subsection{RMS diagram}
\label{rmsdia}

In order to better assess the performance of \texttt{DOHA}, we constructed the RMS diagrams of both the raw and the \texttt{DOHA}-corrected light curves for all stars in our
sample. The resulting diagrams are shown in the left panel of Figure\,\ref{fig5}. The dashed black line indicates the theoretical noise floor curve.

To appreciate the light curve improvement, we can define the \emph{relative RMS improvement} as $1\,-\rm{RMS_{DOHA}}/\rm{RMS_{RAW}}$. This is plotted in the middle panel of 
Figure\,\ref{fig5} against target magnitude. In general, brighter stars show larger improvement, a result of the fact that for brighter stars it is easier to find comparisons with 
very high $R_{tp}$ values (see again Fig.\,\ref{fig4}). We should also note that very large relative improvement factors (>0.8) should be interpreted with some caution. As mentioned 
before (and see also Fig.\,\ref{fig1} and \ref{fig7c}) \texttt{DOHA} performs very well with outliers; part of the improvement can be attributed exactly to outlying points being 
brought to the correct level. For this reason, in the right-hand panel of Figure\,\ref{fig5} we again plot the relative improvement, but this time using the \emph{median
absolute deviation (MAD)}, that is $1\,-\rm{MAD_{DOHA}}/\rm{MAD_{RAW}}$.

Using statistics from the Kepler mission, \cite{howard} estimate that there are 0.0066 transiting hot Jupiters per star, with orbital periods up to 10 days. The deepest, 
ground-observed, transiting exoplanet so far is HATS-6b, with $(R_{P}/R_{\star})^{2} = 0.0323$ \citep{hartman}. From Fig.\,\ref{fig5}, we calculate that in our raw sample, $\sim17\%$ 
of the light curves have the required accuracy to detect such a transit. This percentage increases to $\sim35\%$ after applying \texttt{DOHA}, i.e. there is a factor of 2 improvement.

\subsection{Comparison with SysRem}
\label{srmvdoha}

\begin{figure*}
\includegraphics[width=5.8cm]{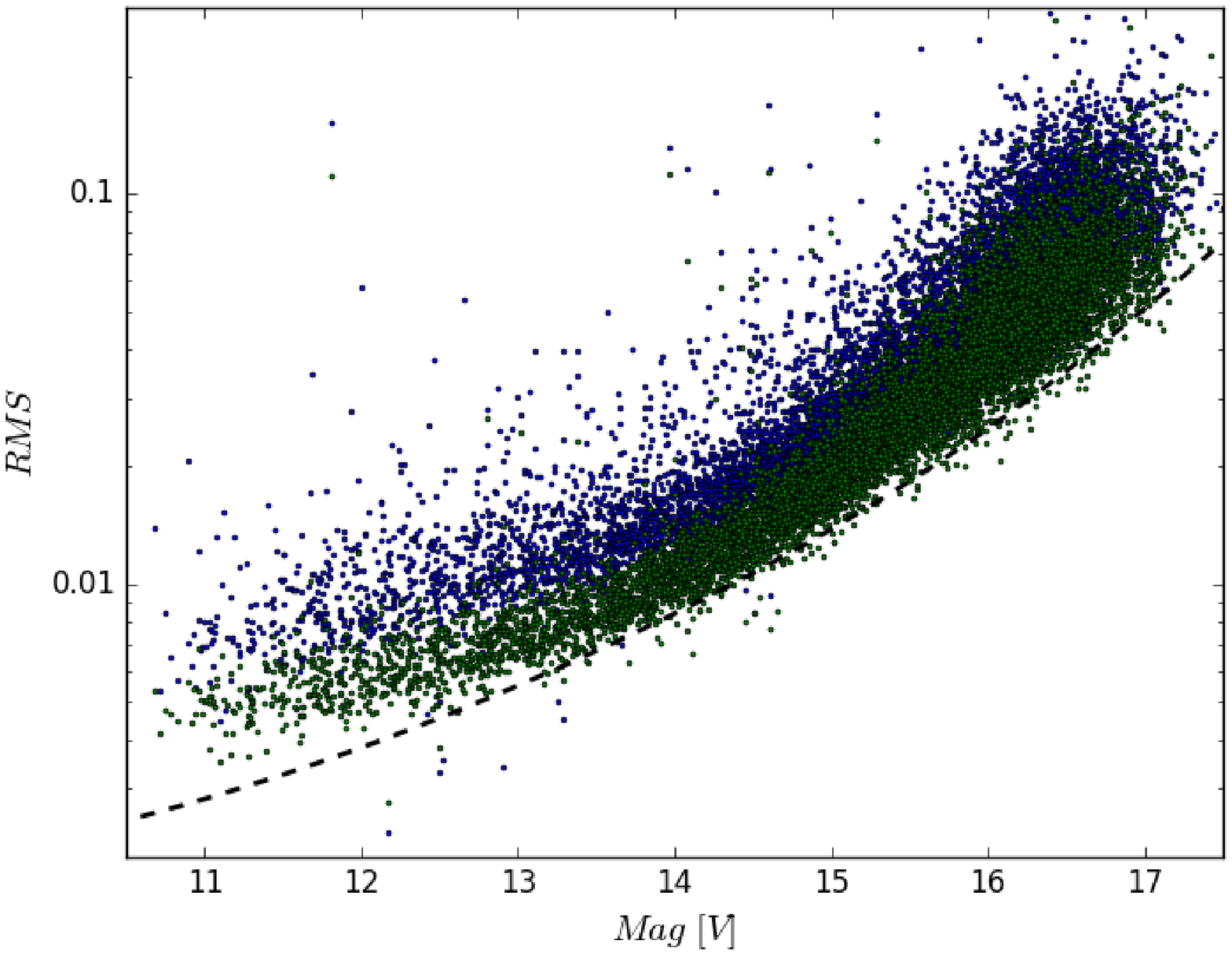} 
\includegraphics[width=5.8cm]{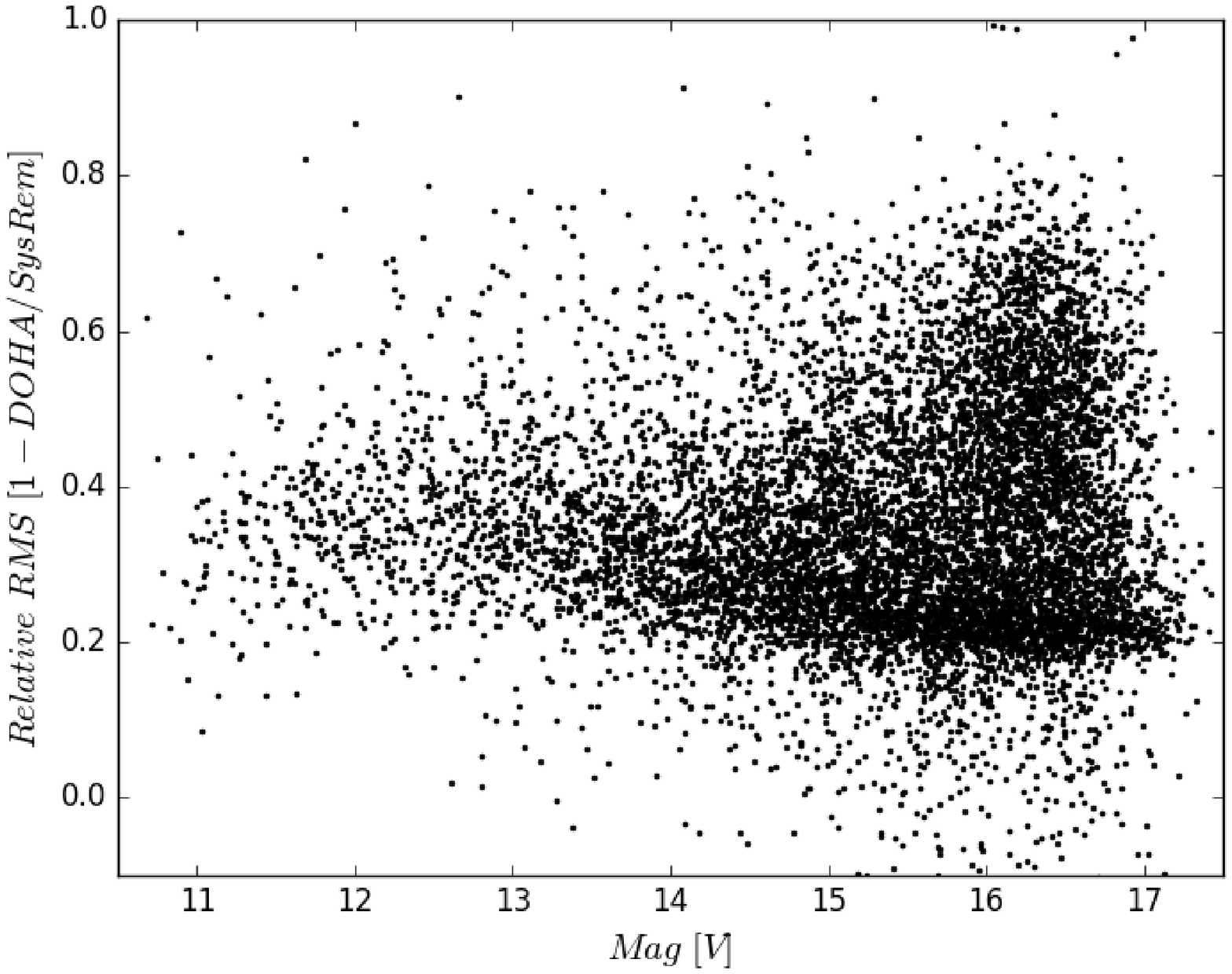} 
\includegraphics[width=5.8cm]{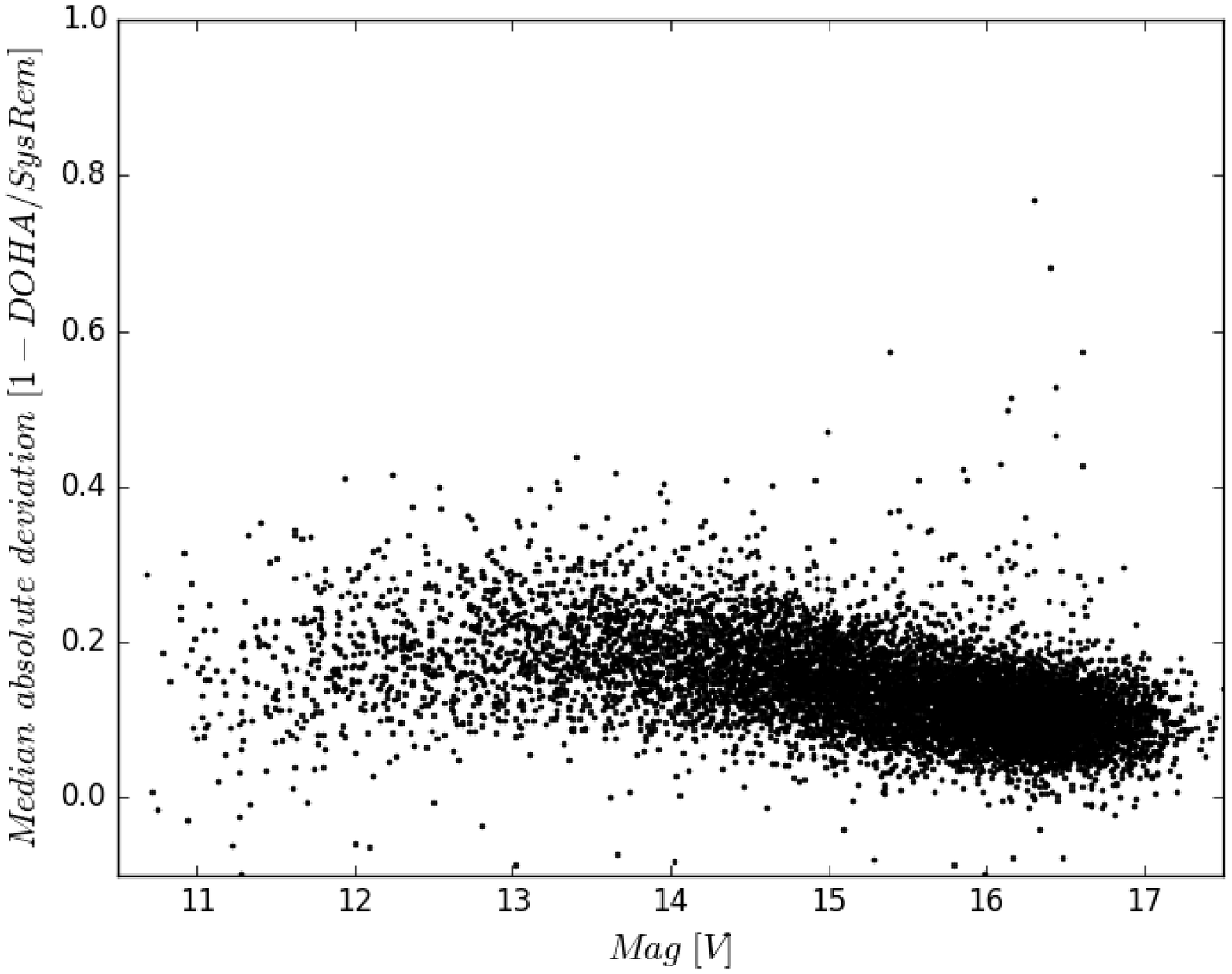} 
 \caption{Same as Fig.\,\ref{fig5}, but this time for \texttt{SysRem} (blue dots) and \texttt{DOHA} (green dots).}
 \label{fig6}
\end{figure*}

As a further performance test, we subjected our sample to detrending using the \texttt{SysRem} algorithm, and compare it with the \texttt{DOHA} results from the previous section. 
Figure\,\ref{fig6} shows the resulting RMS diagrams in the left panel, and the relative RMS and MAD improvements in the middle- and right panel respectively. There is a rather small 
percentage ($\sim$2\% of the total sample) of stars where \texttt{SysRem} yields a better RMS. Also, the same considerations about the treatment of outliers, mentioned previously, 
apply here.

The main difference between \texttt{SysRem} and \texttt{DOHA} is the assumption on the nature of the patterns affecting the data. The implicit assumption of \texttt{SysRem} is 
that the common-mode patterns are dominant and can be expressed as linearly varying components, calculated from the \emph{entire} sample and, furthermore, that these calculated 
components are representative of the entire sample and can therefore be used to correct it. On the other hand, \texttt{DOHA} makes no assumption on the nature or the dominance of the 
patterns (common or uncommon, as described in the introduction) and, moreover, \texttt{DOHA} tries to find the representative components for each star individually, without being 
based on whole sample statistics. To illustrate the point of uncommon dominant patterns, we refer the reader to Fig.\,\ref{fig2} again, where it can be seen that (a) the majority of 
stars actually show very little correlation with the target, (b) practically a quarter of the stars shows, in fact, negative correlation and (c) there are indeed stars which show
high correlation.

\subsection{Individual light curves}
\begin{figure*}
\includegraphics[width=5.3cm]{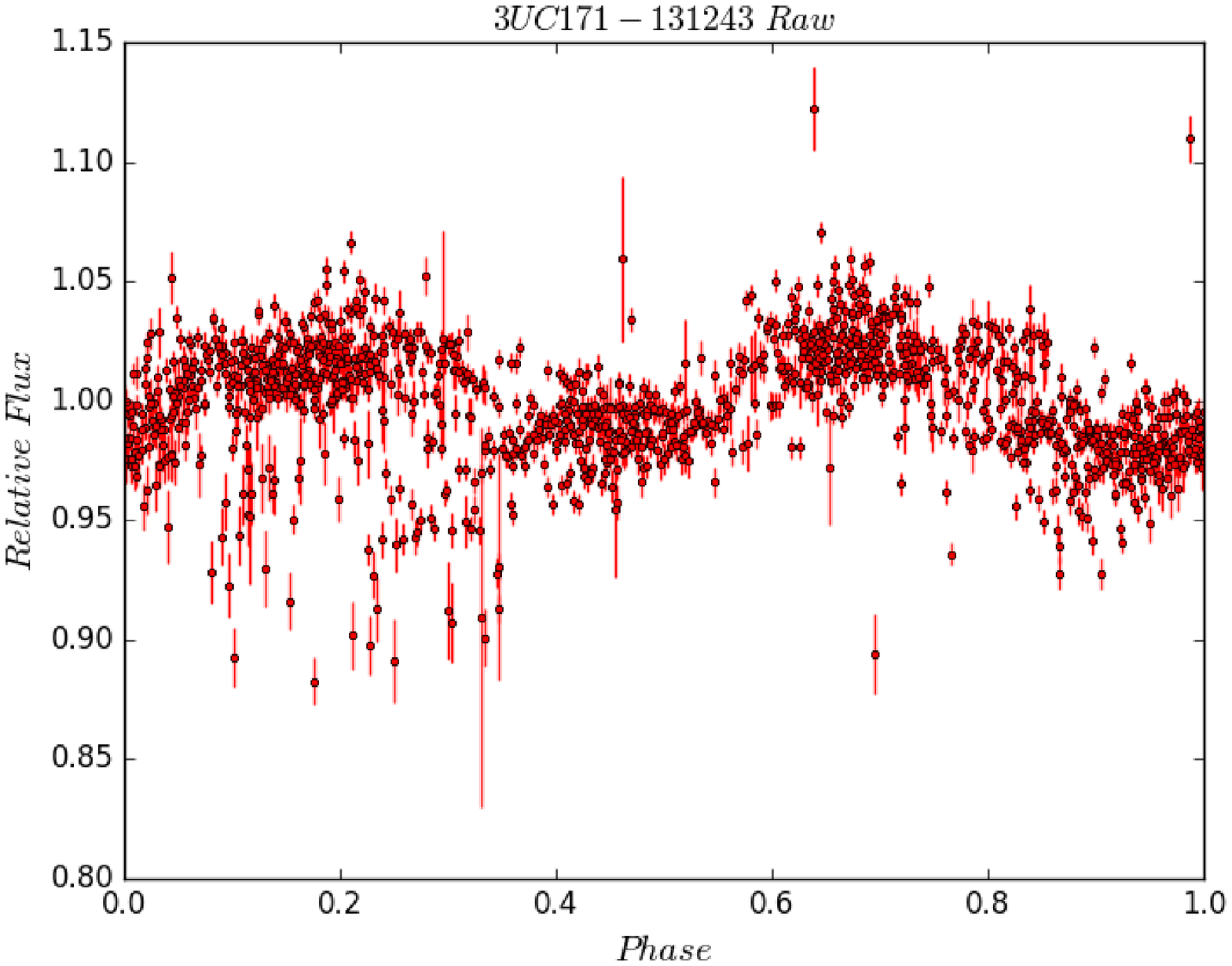}
\includegraphics[width=5.3cm]{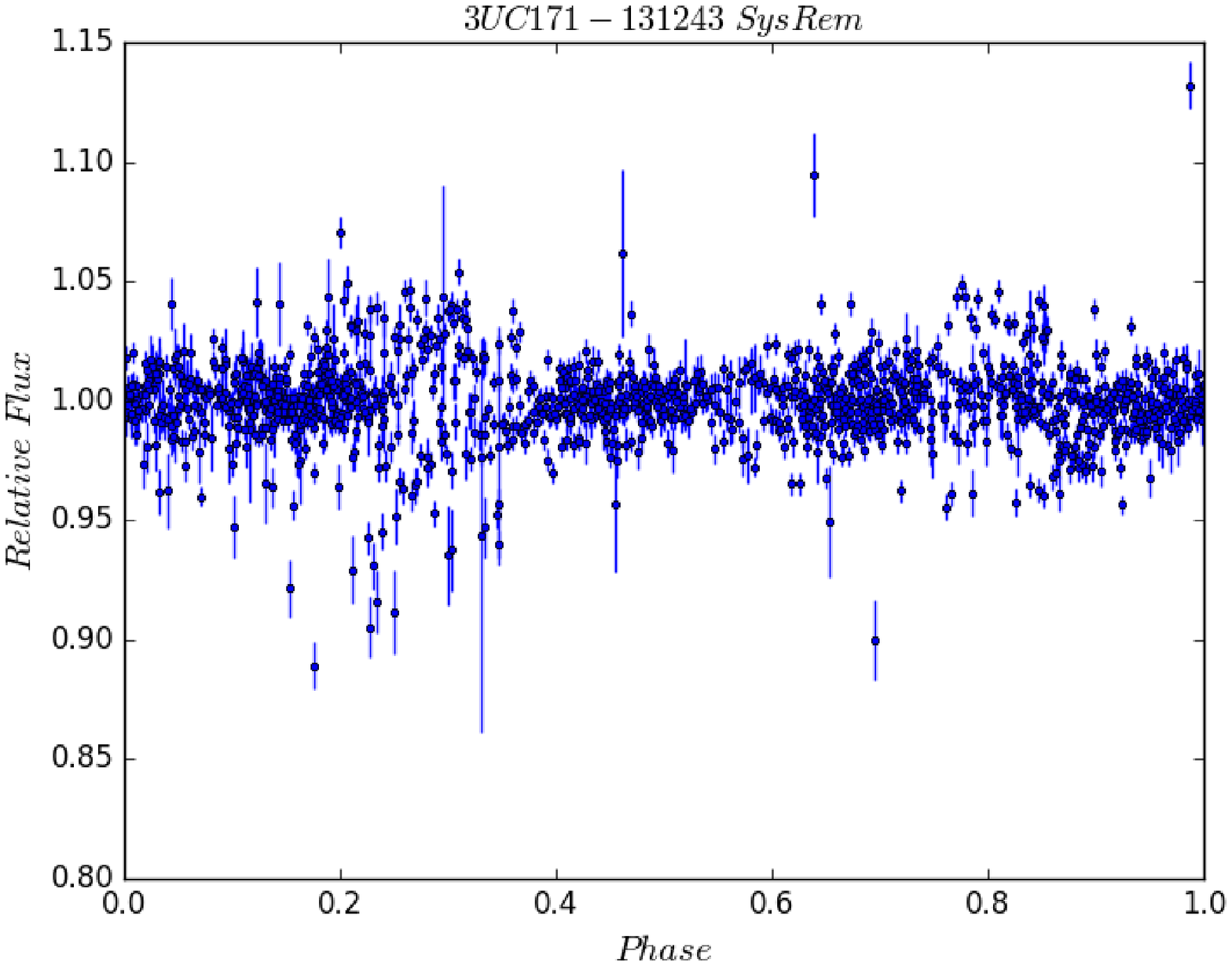}
\includegraphics[width=5.3cm]{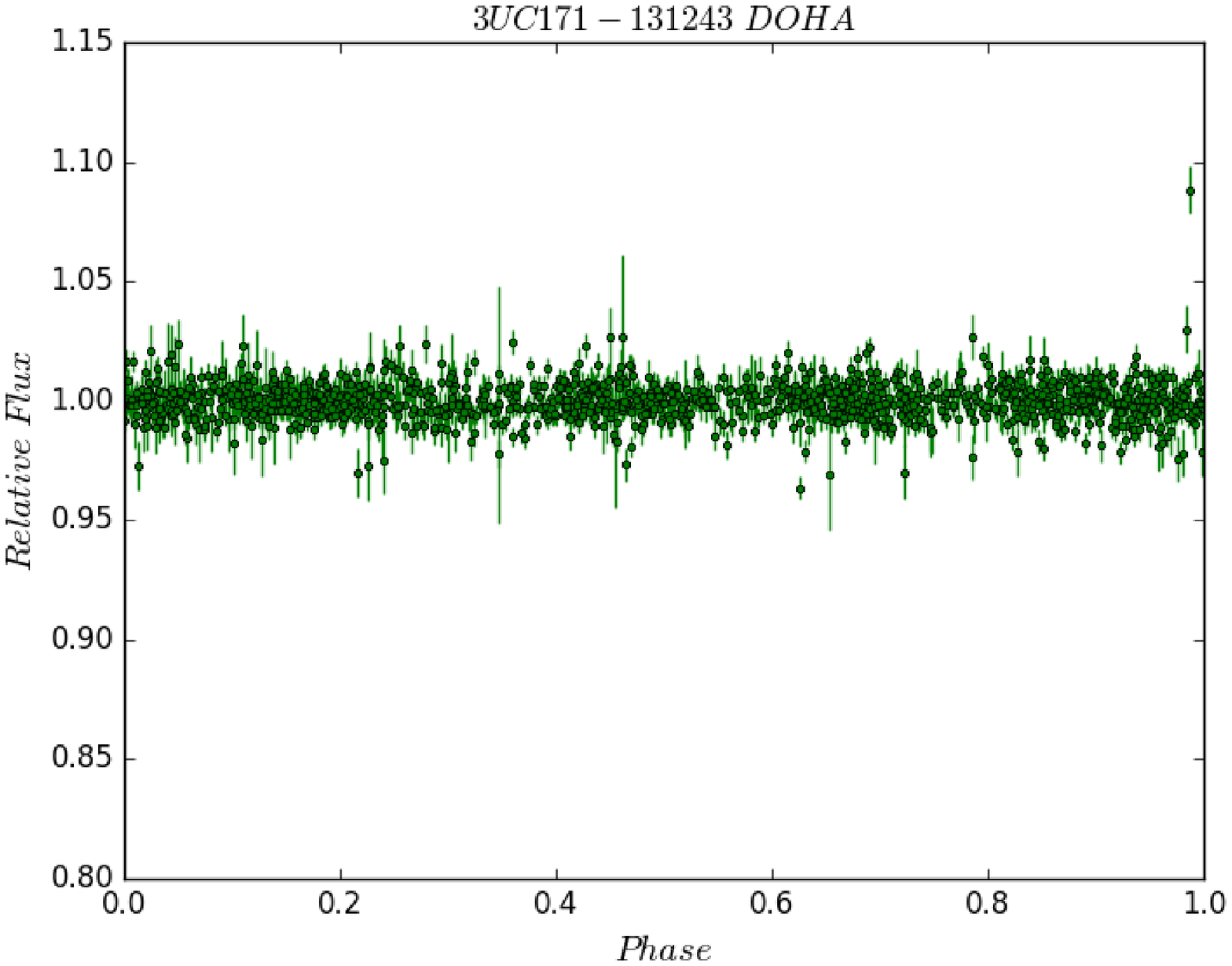}\\
\includegraphics[width=5.3cm]{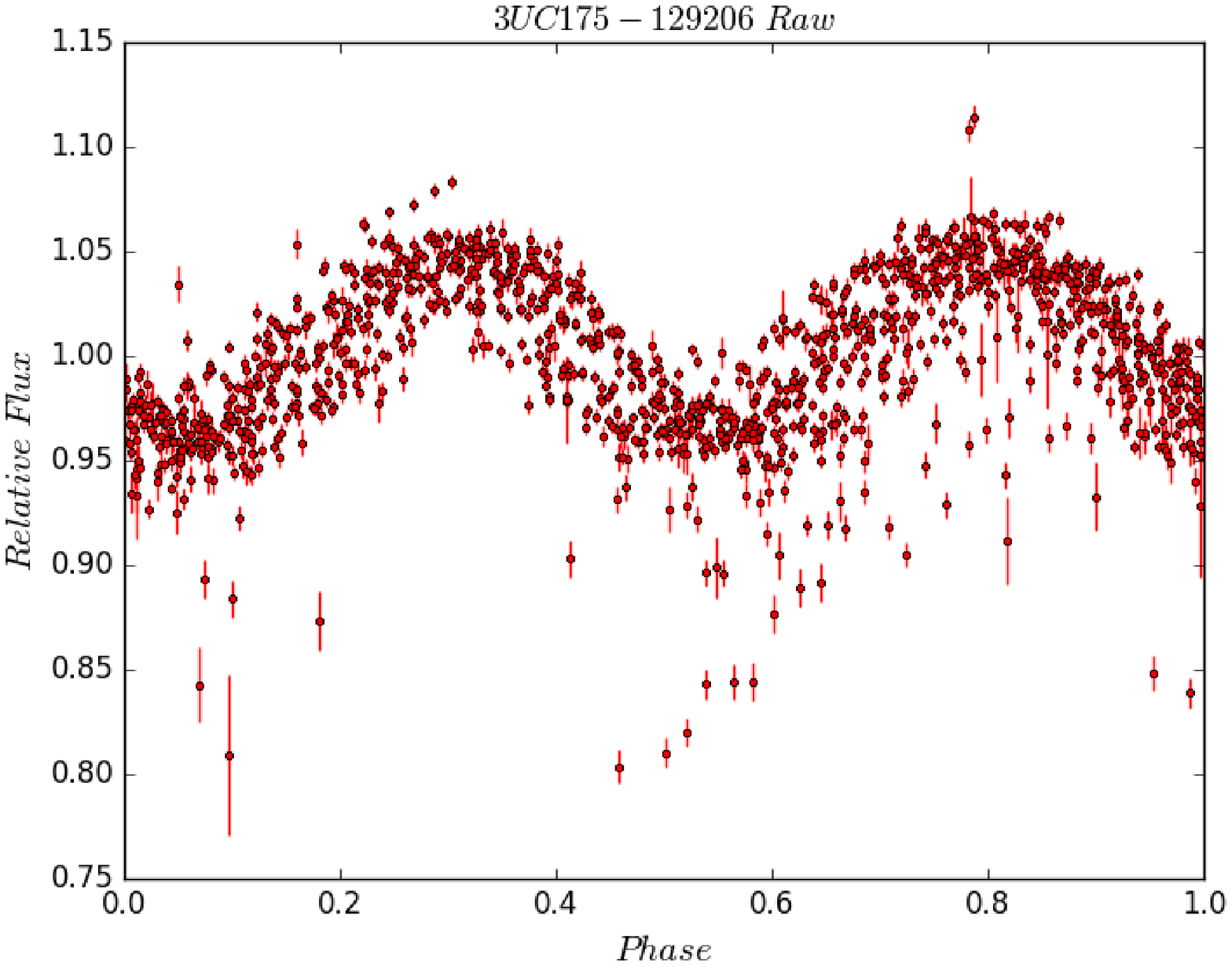}
\includegraphics[width=5.3cm]{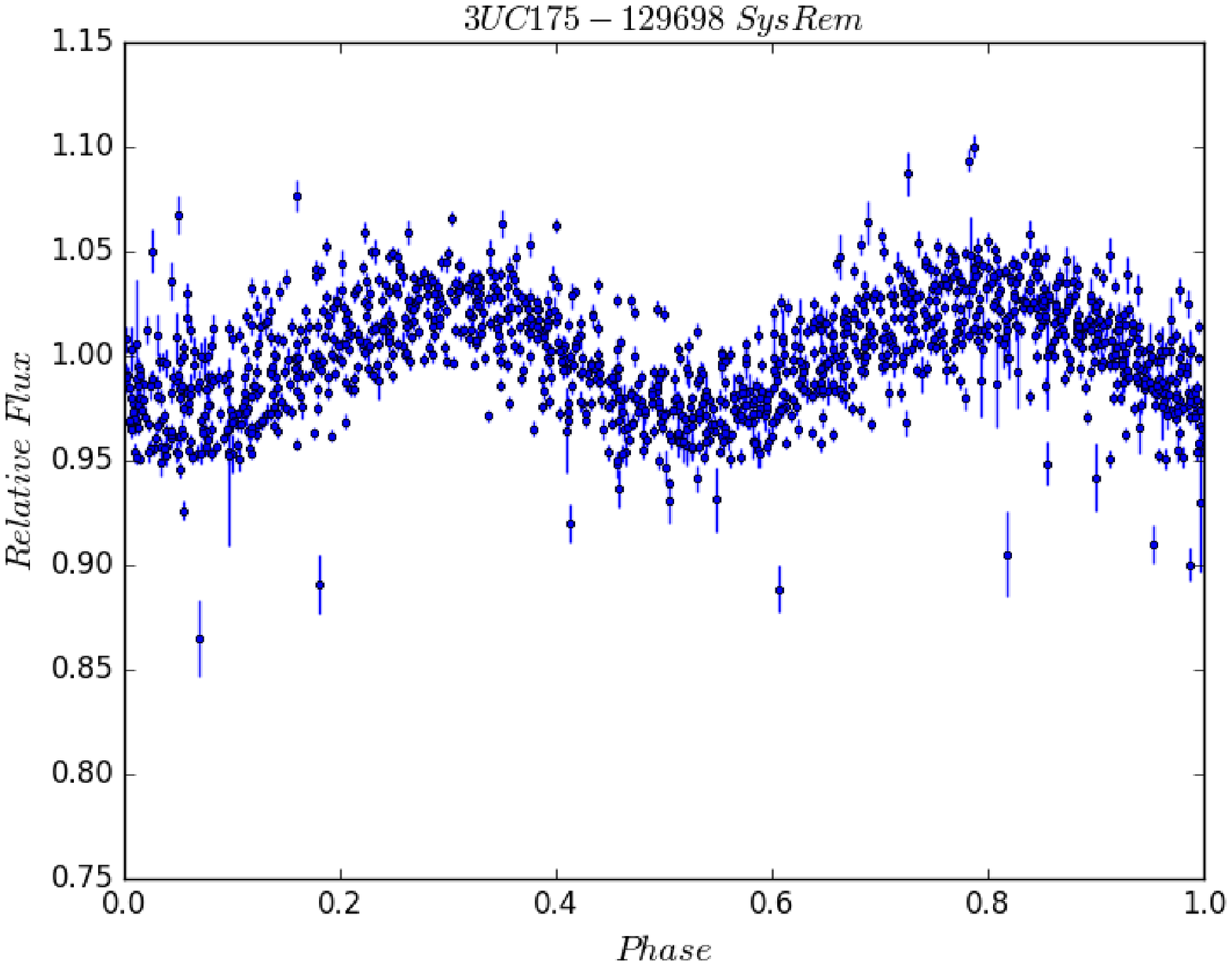}
\includegraphics[width=5.3cm]{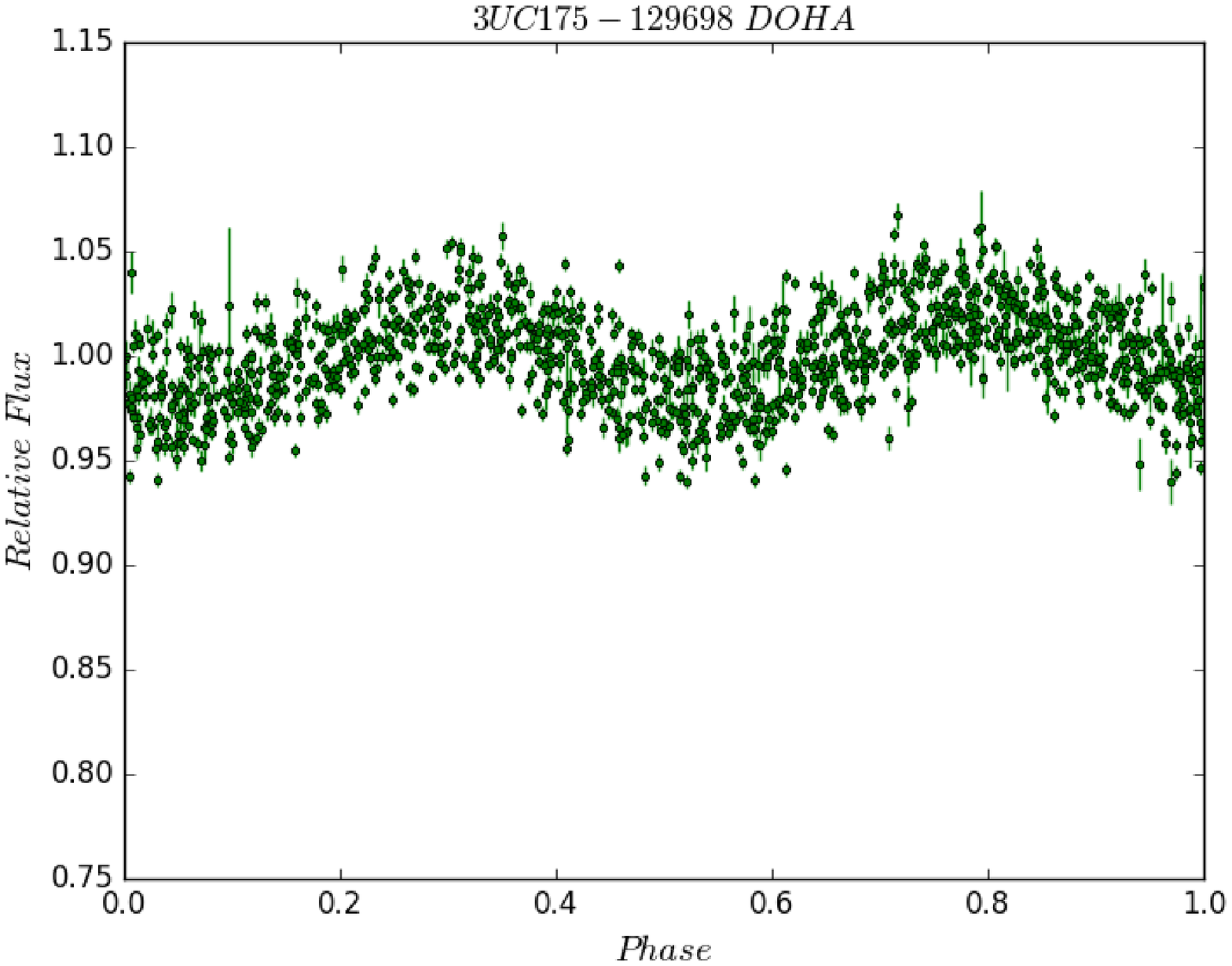}\\
\includegraphics[width=5.3cm]{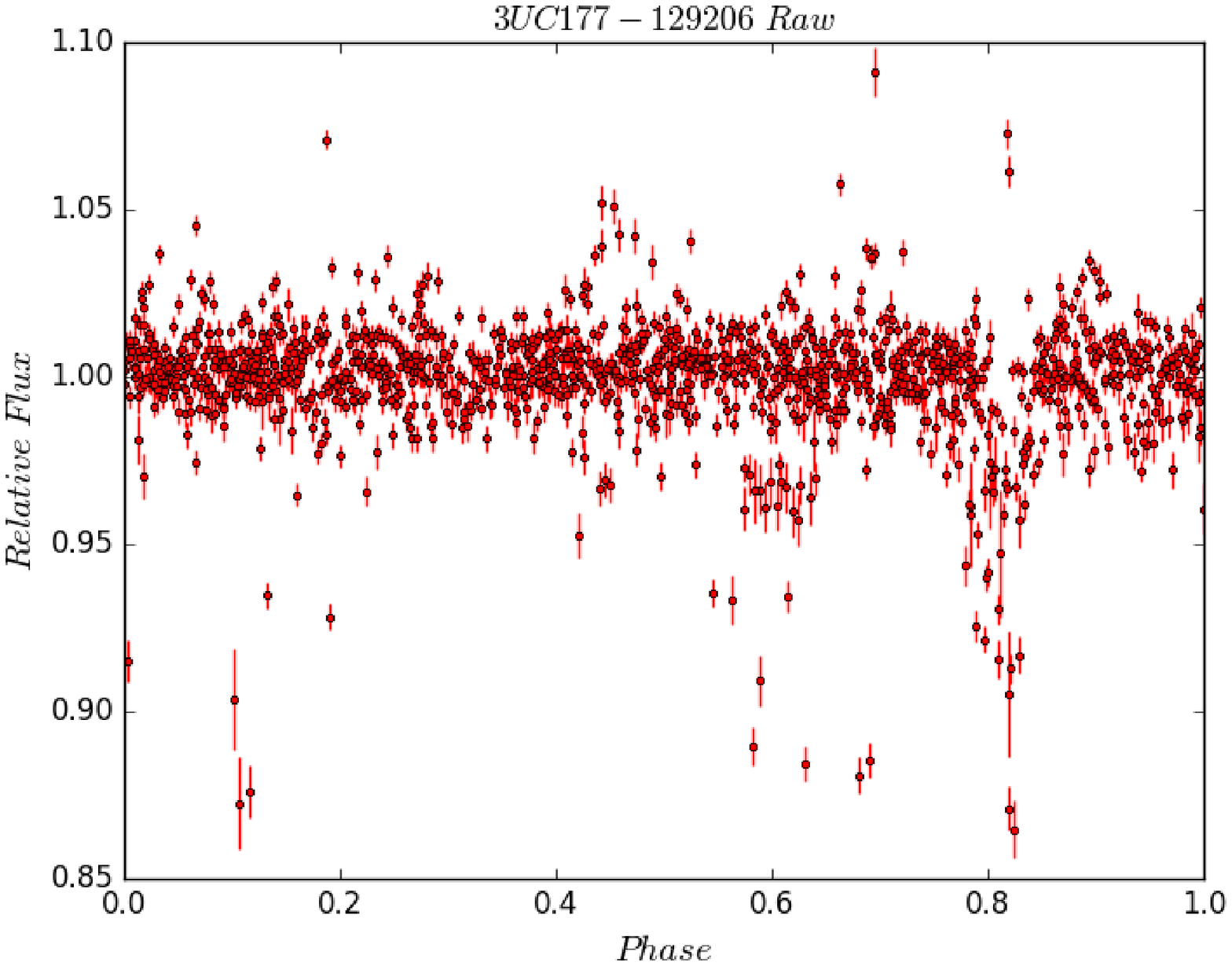}
\includegraphics[width=5.3cm]{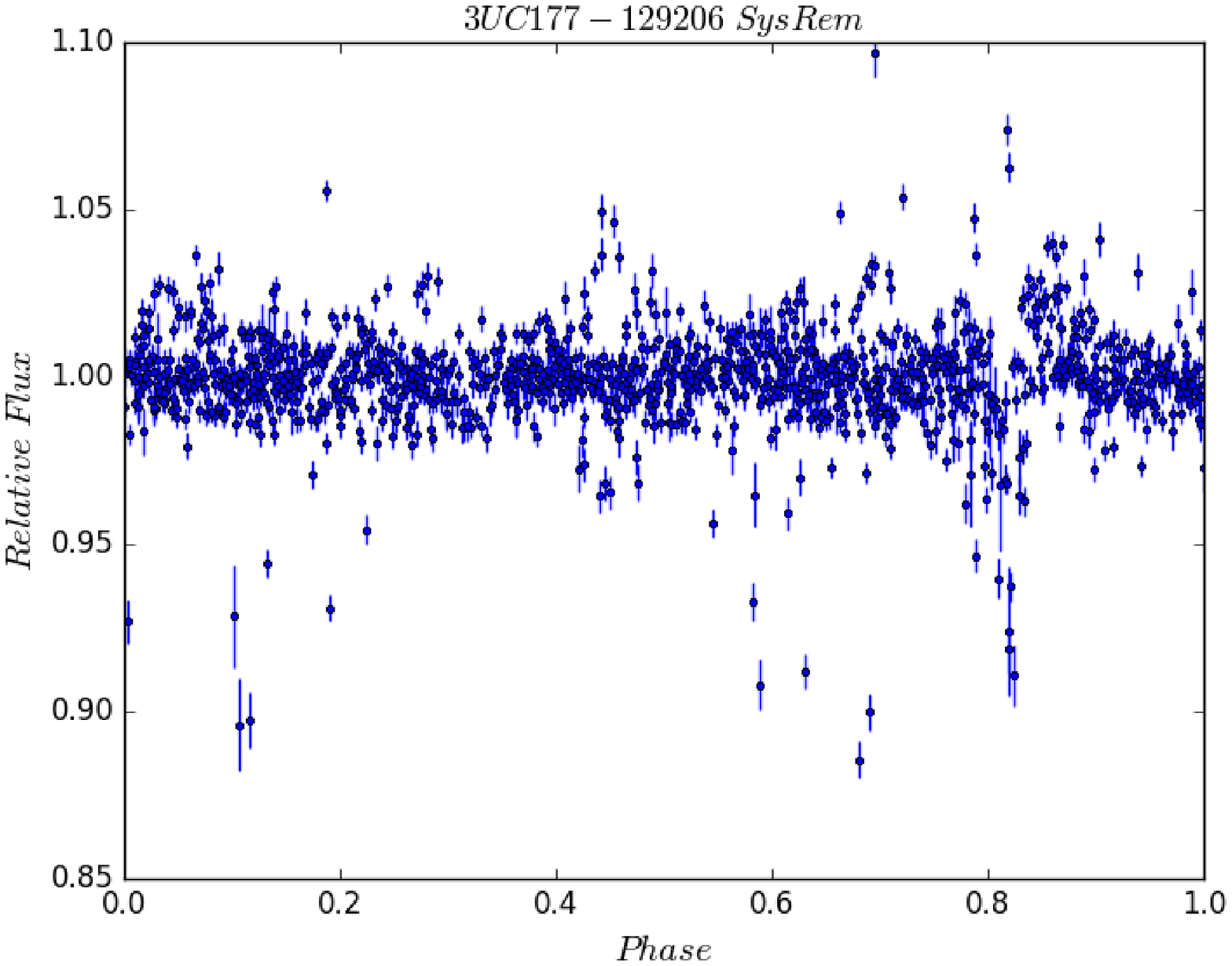}
\includegraphics[width=5.3cm]{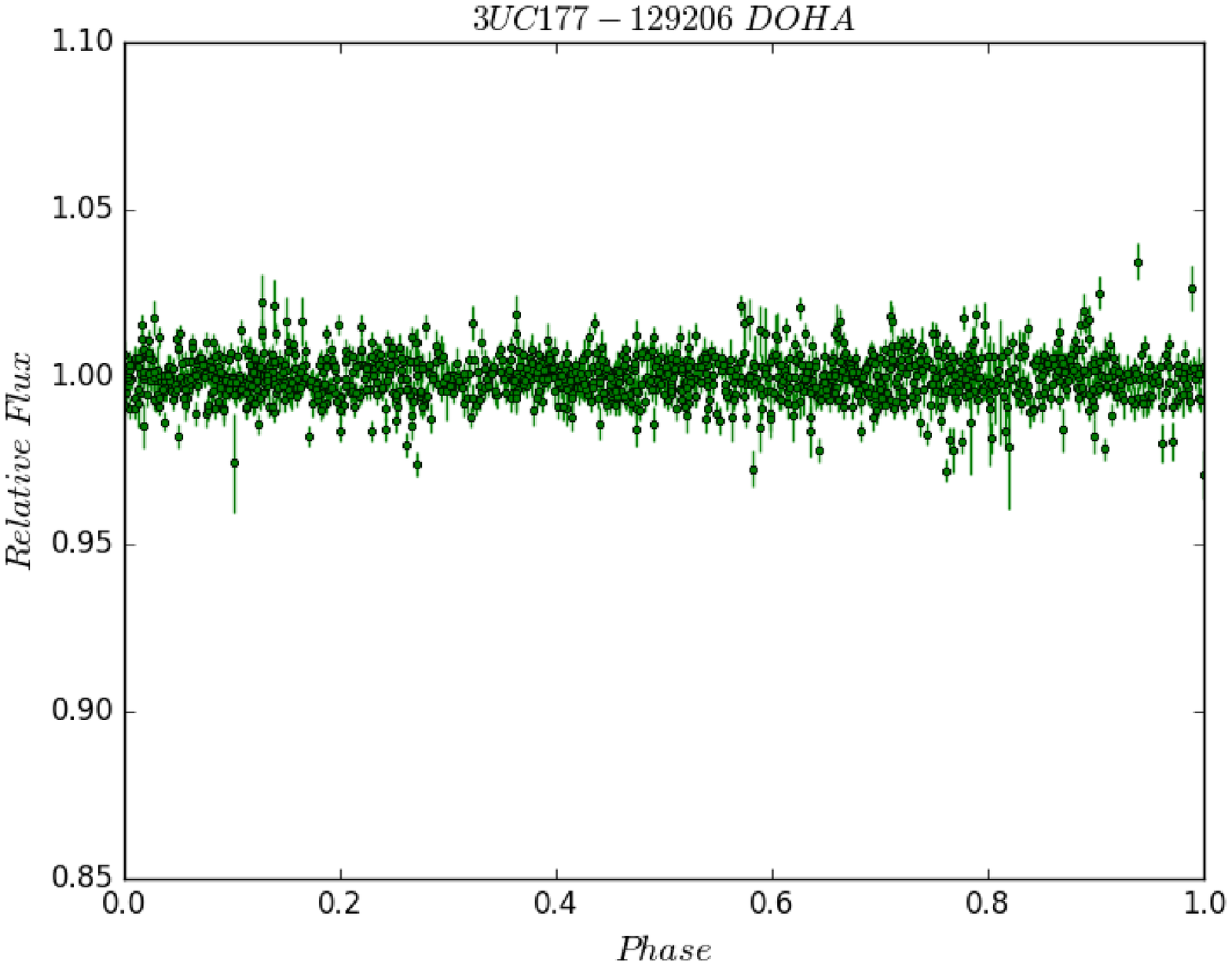}\\
\caption{The light curves of (top to bottom) 3UC171-131243, 3UC175-129698 and 3UC177-129206. \textbf{Left:} raw light curve; \textbf{Middle:} light curve corrected with 
\texttt{SysRem}; \textbf{Right:} light curve corrected with \texttt{DOHA}. The data were phase-folded using the period obtained from running \texttt{BLS}. See Table 1 for more 
information on the stars and the periods.}
\label{fig7c}
\end{figure*}

\subsubsection{General examples}
To illustrate the performance of \texttt{DOHA} more accurately, we present here three individual light curves selected from our sample. These light curves are representative examples 
of the patterns affecting our data. Table\,1 gathers some basic information on these three stars. The reported periods come after running the ``Box Least Squares'' (\texttt{BLS}) 
algorithm of \citet{kovacs1}.

\begin{itemize}
 \item[$\bullet$] 3UC171-131243 (Fig.\,\ref{fig7c}, top panels): this is a constant star, but systematics introduce non-real variability, which is picked-up by the BLS search. \texttt{DOHA}, 
 not only corrects the systematics creating the variability, but also corrects almost all the outliers between phase $0.1 < \phi < 0.3$. \\
 \item[$\bullet$] 3UC175-129698 (Fig.\,\ref{fig7c} middle panels): this is a typical short period variable star, ($P\,=\,0.391388\,[\mathrm{d}]$) Note that the amplitude of the variability 
 remains unaffected after applying \texttt{DOHA}. Furthermore, similar to the previous example, the algorithm manages to correct almost all the outlying points. \\
 \item[$\bullet$] 3UC177-129206 (Fig.\,\ref{fig7c} bottom panels): the light curve of this system seems to contain a ``transit-like'' signature at phase $\phi\sim 0.8$. The corrected light 
 curve is much cleaner without outliers, but, most importantly, the ``transit-like'' signal disappears. This is a case where \texttt{DOHA} successfully corrects a false-positive 
 identification. The fact that this star is, indeed, constant is supported by radial velocity measurements which show no RV variations, to a level of $K<40$ $\mathrm{m sec^{-1}}$, 
 corresponding to a mass of smaller than $0.2M_{J}$.
\end{itemize}

\begin{table}
\centering
\label{tab:char}
 \caption{Basic information for the three example stars. }
\begin{tabular}{ccccc}
\hline
UNSO4 ID & RA & Dec & Mag & BLS period [d] \\
\hline
3UC171-131243 & 13:57:16.09 & -04:31:51.10 & 13.4 & 1.966354 \\
3UC175-129698 & 13:52:05.31 & -02:45:32.70 & 13.0 & 0.391308 \\
3UC177-129206 & 13:54:36.36 & -01:42:11.80 & 12.8 & 1.596665  \\
\hline
 \end{tabular}
\end{table}
\noindent

\begin{figure*}
\includegraphics[width=8cm]{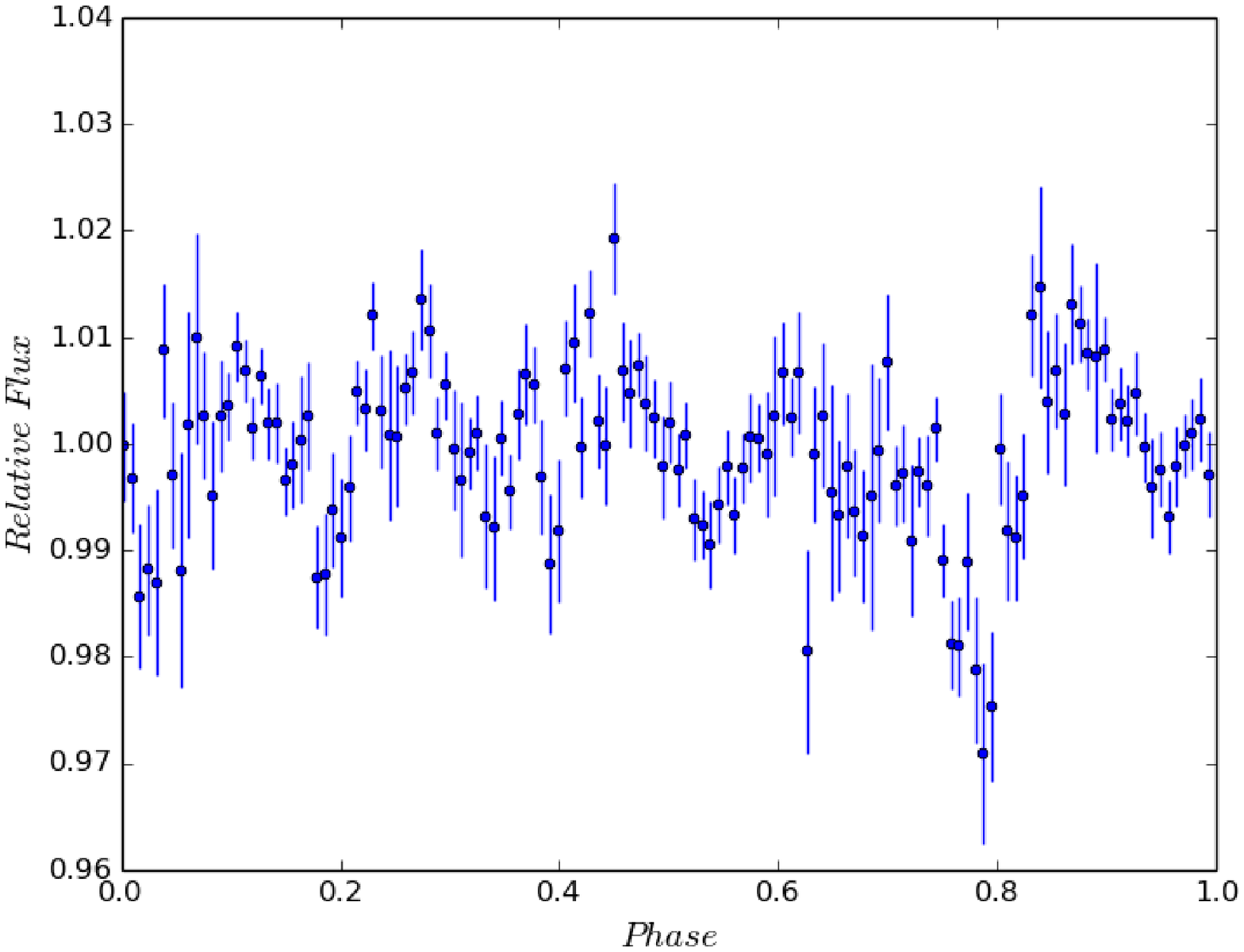}
\includegraphics[width=8cm]{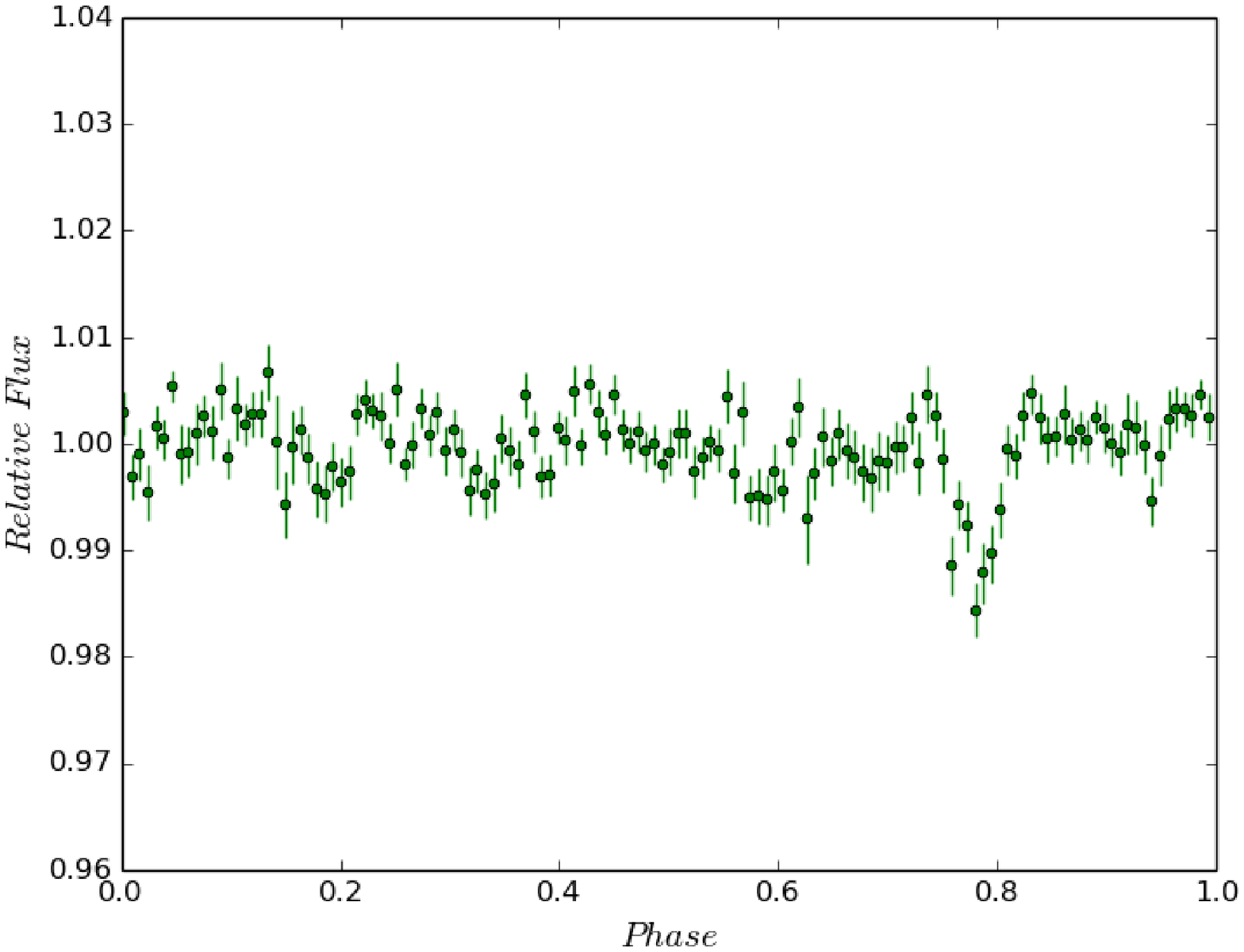}\\
 \caption{\textbf{WASP-1b} phase-folded and binned light curves using \texttt{SysRem} (left panel) and \texttt{DOHA} (right panel) for detrending.}
\label{fig8a}
\end{figure*}

\begin{figure*}
\includegraphics[width=8cm]{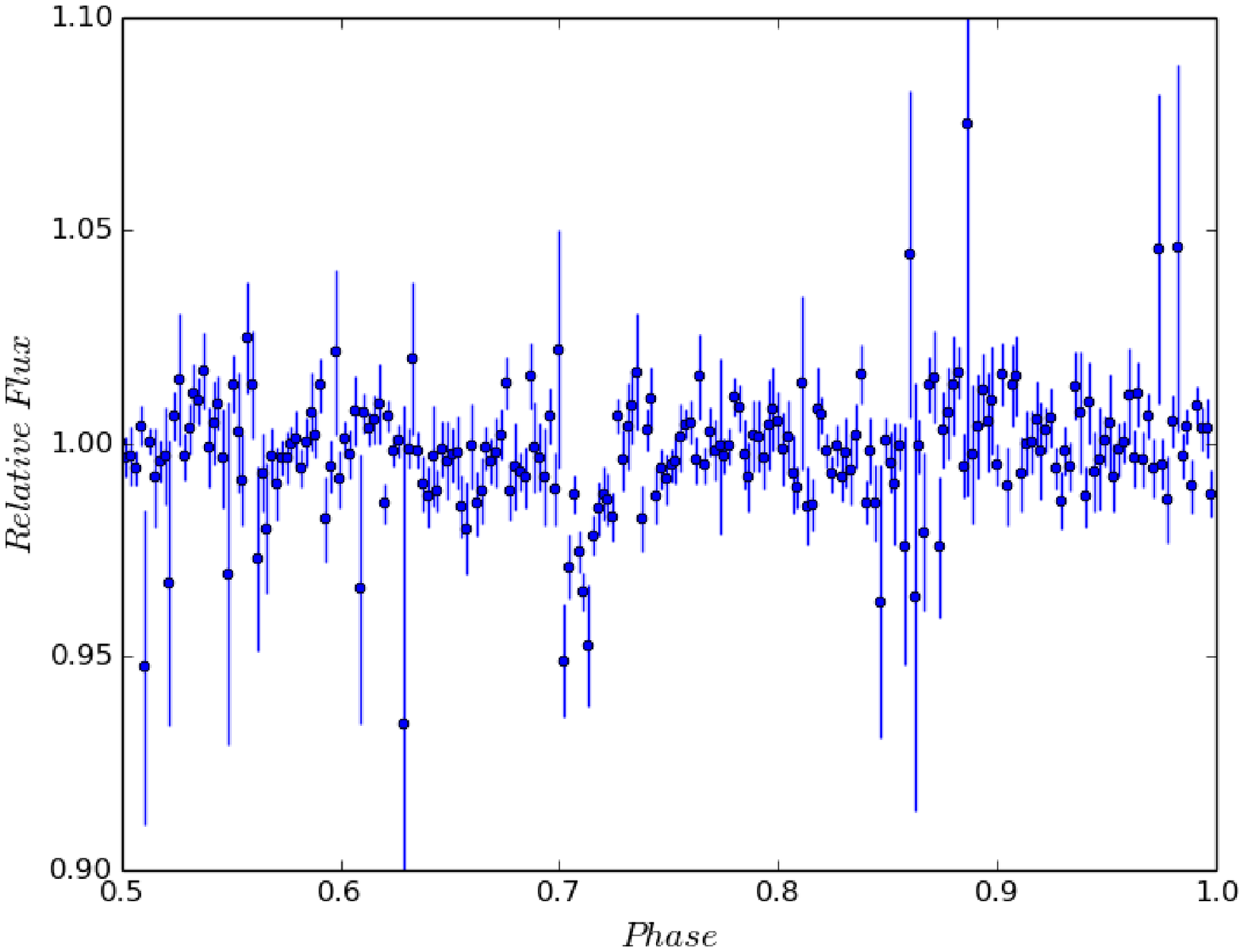}
\includegraphics[width=8cm]{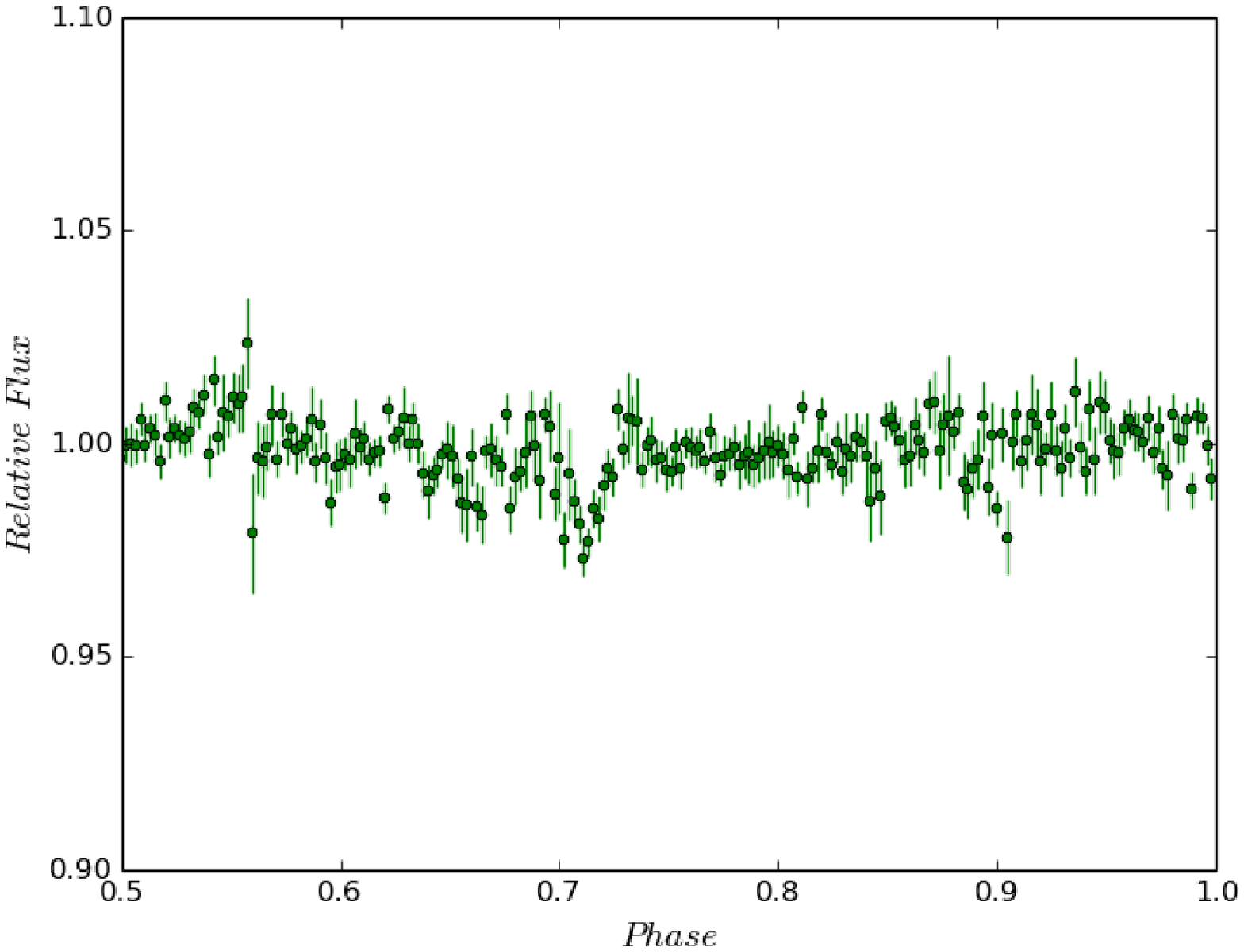}\\
 \caption{Same as Fig.\,\ref{fig8a}, but for \textbf{HAT-P-19b}.}
\label{fig8b}
\end{figure*}

\begin{figure*}
\includegraphics[width=8cm]{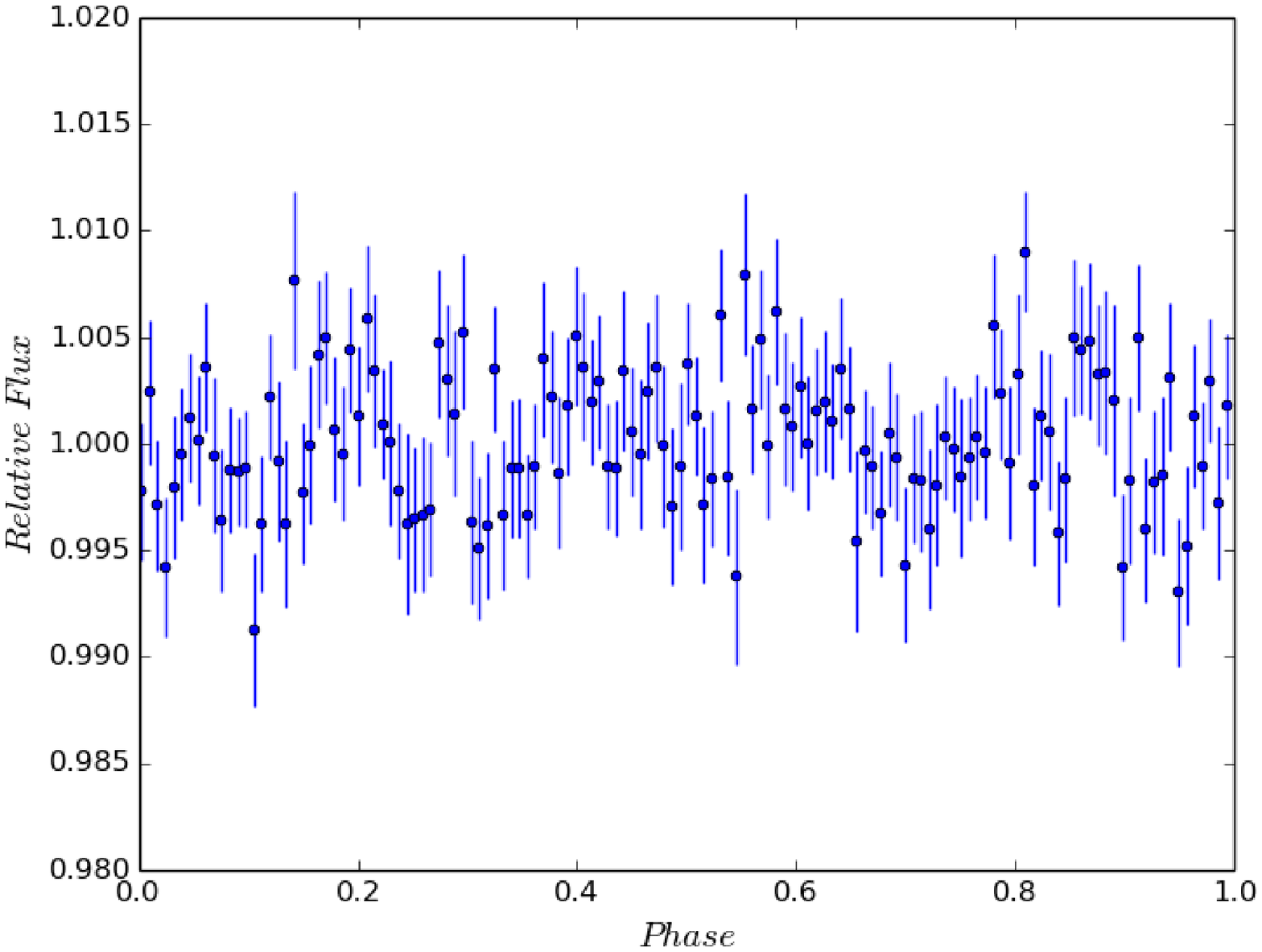}
\includegraphics[width=8cm]{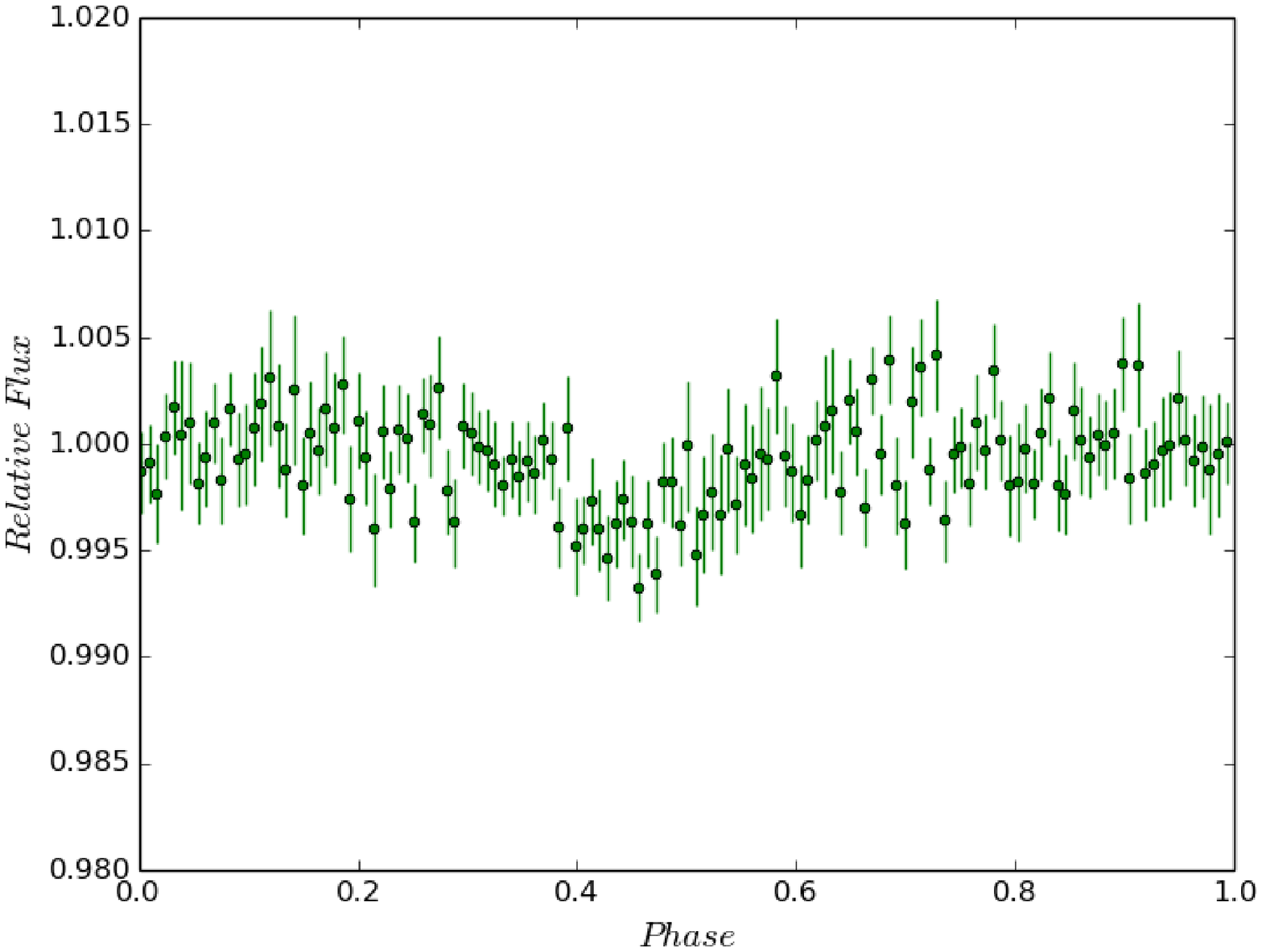}
 \caption{Same as Fig.\,\ref{fig8a} \& \ref{fig8b}, but for \textbf{KELT-1b}.}
\label{fig8c}
\end{figure*}

\subsubsection{Transiting light curves}

A further test was to assess the performance of \texttt{DOHA} on known transiting exoplanets that have been observed with the QES. We have selected data that actually contain three
known planets in the same field\footnote{This field is different than the one described in Sec.\,\ref{sec:sample}.}: WASP-1b \citep{cameron} with period $P_{W}=2.5199464$ days and 
magnitude $mag_{W}=11.63$; HAT-P-19b \citep{hartman2} with period $P_{H}=4.008784$ days and magnitude $mag_{H}=12.9$; and KELT-1b \citep{siverd} with period $P_{K}= 1.217514$ days and 
magnitude $mag_{K}=10.7$. KELT-1b is a very massive object (27$M_{J}$), but because it is orbiting a mid-F type star, the depth of the transit is very small (0.6\%), making it an ideal
target for our test.

We first detrended the raw data using \texttt{SysRem} and subsequently ran \texttt{BLS} on the resulting light curves. \texttt{BLS} successfully detected WASP-1b and HAT-P-19b, but 
failed to detect KELT-1b. We then, repeated the process, only this time we corrected the raw light curves using \texttt{DOHA}. In this case, all three planets were successfully 
detected by the \texttt{BLS}, with the correct parameters for orbital period and transit depth. Figures\,\ref{fig8a}-\ref{fig8c} show the \texttt{SysRem} (left panels) versus 
\texttt{DOHA} (right panels) phase-folded and binned light curves of the three planets.

\section{Signal detection algorithms}
\label{sec:sigdet}

As a final test, we investigated the effect of \texttt{DOHA} on the performance of signal detection algorithms. For comparative purposes, we also ran the same tests using 
the \texttt{SysRem}-detrended light curves, as presented in Sec.\,\ref{srmvdoha}.

The tests were conducted as follows: we injected simulated transit signals, generated using the \cite{pal} model, in all the 9374 raw light curves of our sample 
(Sec.\,\ref{sec:sample}). For all transits, the stellar and planetary parameters were kept fixed to $R_{\star}$ = 1.0 $R_{\odot}$, $M_{\star}$ = 1.0 $M_{\odot}$ and $R_{P}$ = 1.0 
$R_{J}$, while the orbital period was randomly chosen from a uniform distribution, with $1\,[\mathrm{d}]\,<\,P_{orb}\,<\,5\,[\mathrm{d}]$. This combination of stellar and planetary 
parameters was chosen to ensure a large number of detections for statistical purposes.

Subsequently, the raw light curves (now including the transit signals) were subjected to correction using both \texttt{SysRem} and \texttt{DOHA}. In each corrected set, we 
searched for transits using two separate signal detection algorithms: \texttt{BLS} and \texttt{SiDRA} \citep{mislis1}. We note that the test was not designed to compare \texttt{BLS} 
with \texttt{SiDRA}, only to assess how the probability of detecting a transit, using each detection algorithm, changes after applying \texttt{DOHA}.

\begin{figure}
\includegraphics[width=8cm]{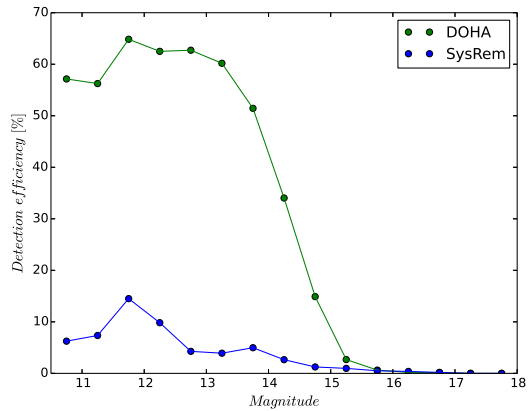} 
 \caption{BLS detection efficiency using SysRem (blue curve) and \texttt{DOHA} (green curve) in each 0.5mag-wide magnitude bin.}
\label{fig9}
\end{figure}

The combination of \texttt{SysRem}+\texttt{BLS} yielded 149 transits (1.6\% of the total), whereas \texttt{DOHA}+\texttt{BLS} successfully identified 1226 transits (13.1\% of the 
total). To have a clearer view, we divided our sample in 0.5mag-wide bins and in Figure\,\ref{fig9} we plot the \texttt{BLS} detection efficiency in each magnitude bin. If we now
restrict the considered magnitude range to $mag<14$ (the working magnitude range of the QES survey), then \texttt{BLS} correctly identifies 6.2\% of the transits, using 
\texttt{SysRem}; and 58\% of the transits, using \texttt{DOHA}. 

For the test with \texttt{SiDRA}\footnote{\texttt{SiDRA} is an entropy-based, random forest classification algorithm, and does not yield physical parameters, such as the
orbital period.}, we imposed a strict 70\% confidence cut-off \citep[see][for details]{mislis1}. At this level, \texttt{SiDRA} classified 505 systems as definite planets using the 
\texttt{SysRem} light curves (5.4\% of the total); and 938 systems (10.0\% of the total) using the \texttt{DOHA}-corrected light curves. If we again restrict the magnitude range, as 
before, then \texttt{SiDRA} returns 7\% of the total number of planets, using \texttt{SysRem}; and 20\% using \texttt{DOHA}. 

As a by-product, using the \texttt{DOHA} light curves, \texttt{SiDRA} correctly identified three (already known) variables in the field (2 RR\,Lyr and 1 W\,Uma) which
have been missed in a variable search using \texttt{SysRem}-detrended light curves.
 
It is evident that \texttt{DOHA} significantly increases the chances of finding transiting planets, regardless of the signal detection algorithm employed. 

\section{Conclusions}
\label{sec:conc}
In this paper we have presented \texttt{DOHA}, a new algorithm for correcting light curves obtained with large-scale, ground-based photometric surveys, with an emphasis on 
those of transiting exoplanets. Adopting the reasoning of a comment made by the referee during the review process, we denote \texttt{DOHA} as a \emph{cotrending}, rather than a
detrending algorithm.

\texttt{DOHA} is based on the standard differential photometry technique of correcting a target light curve using a master comparison light curve constructed from suitable,
individual comparison stars. The success of \texttt{DOHA} lies in its ability to optimise the way in which suitable comparison stars are selected for \emph{each} target 
\emph{separately}. \texttt{DOHA} looks for and corrects common-mode patterns shared by the target and a ``base'' of comparison stars (constituting a small subset of all stars in the 
field), which is built after a two-step correlation search; the first accounting for long-term trends, the second for intra-night variations. In short, \texttt{DOHA} exploits the
defining characteristic of systematics, that is their manifestation as common-mode behaviour of the data, without making any assumptions on their nature and prevalence. As such, 
\texttt{DOHA} is able to correct data trends and patterns regardless of their commonness and/or individual contribution to the variations in the sample. Our algorithm can either be 
used as stand-alone on raw light curves, or as a compliment to detrending algorithms, correcting for residual uncommon patterns.

To test and assess the performance of \texttt{DOHA}, we have used $\sim$9500 light curves from the QES transiting survey. The results show that \texttt{DOHA} is able to improve the 
light curve RMS by a factor of 2, doubling the probability of detecting a transit signal. Results also indicate that \texttt{DOHA} is particularly efficient on bright
stars.

Finally, by adding simulated transits in all of our sample light curves, we showed that, using \texttt{DOHA} combined with two separate signal detection algorithms, the 
number of successful detections can increase considerably.

\section*{Acknowledgments}

We would like to thank the anonymous referee for a prompt and useful report. This publication was made possible by NPRP grant $\sharp$ X-019-1-006 from the Qatar National Research 
Fund (a member of Qatar Foundation). The statements made herein are solely the responsibility of the author.

\bsp

\label{lastpage}
\end{document}